\documentclass[11pt]{article}
\usepackage[colorlinks=true, citecolor=blue, urlcolor=blue, linkcolor=blue, breaklinks=true, pdfpagelabels=false]{hyperref}

\usepackage{hyperref}

\usepackage{amsmath,amsfonts}
\usepackage{graphicx}
\usepackage{epsfig,epsf}
\usepackage{bm}

\def\be {\begin{equation}}
\def\ee {\end{equation}}
\def\beq {\begin{equation}}
\def\eeq {\end{equation}}
\def\bea {\begin{eqnarray}}
\def\eea {\end{eqnarray}}

\def\bra {\langle}
\def\ket {\rangle}

\def\beq{\begin{equation}}
\def\eeq{\end{equation}}
\def\barr{\begin{array}}
\def\earr{\end{array}}

\def\opcit(#1){ {\em op. cit.}, #1}

\def\issue(#1,#2,#3){#1, #2 (#3)} 

\def\APP(#1,#2,#3){Acta Phys.\ Polon.\ \issue(#1,#2,#3)}
\def\ARNPS(#1,#2,#3){Ann.\ Rev.\ Nucl.\ Part.\ Sci.\ \issue(#1,#2,#3)}
\def\CPC(#1,#2,#3){Comp.\ Phys.\ Comm.\ \issue(#1,#2,#3)}
\def\CIP(#1,#2,#3){Comput.\ Phys.\ \issue(#1,#2,#3)}
\def\EPJC(#1,#2,#3){Eur.\ Phys.\ J.\ C\ \issue(#1,#2,#3)}
\def\EPJD(#1,#2,#3){Eur.\ Phys.\ J. Direct\ C\ \issue(#1,#2,#3)}
\def\IEEETNS(#1,#2,#3){IEEE Trans.\ Nucl.\ Sci.\ \issue(#1,#2,#3)}
\def\IJMP(#1,#2,#3){Int.\ J.\ Mod.\ Phys. \issue(#1,#2,#3)}
\def\JHEP(#1,#2,#3){J.\ High Energy Physics \issue(#1,#2,#3)}
\def\JPG(#1,#2,#3){J.\ Phys.\ G \issue(#1,#2,#3)}
\def\MPL(#1,#2,#3){Mod.\ Phys.\ Lett.\ \issue(#1,#2,#3)}
\def\NP(#1,#2,#3){Nucl.\ Phys.\ \issue(#1,#2,#3)}
\def\NIM(#1,#2,#3){Nucl.\ Instrum.\ Meth.\ \issue(#1,#2,#3)}
\def\PL(#1,#2,#3){Phys.\ Lett.\ \issue(#1,#2,#3)}
\def\PRD(#1,#2,#3){Phys.\ Rev.\ D \issue(#1,#2,#3)}
\def\PRL(#1,#2,#3){Phys.\ Rev.\ Lett.\ \issue(#1,#2,#3)}
\def\SJNP(#1,#2,#3){Sov.\ J. Nucl.\ Phys.\ \issue(#1,#2,#3)}
\def\ZPC(#1,#2,#3){Zeit.\ Phys.\ C \issue(#1,#2,#3)}

\def\equationautorefname~#1\null{Eq.\,(#1)\null}
\def\pageautorefname\nobreakspace{p.}

\makeatletter\renewcommand{\p@subsection}{\thesection.}\makeatother%
 
\begin{document}

\renewcommand*{\thefootnote}{\fnsymbol{footnote}}


\begin{center}
{\Large\bf{Potential of a singlet scalar enhanced Standard Model}}


\vspace{5mm}

{\bf Swagata Ghosh},$^{a,}$\footnote{swgtghsh54@gmail.com}
{\bf Anirban Kundu},$^{a,}$\footnote{anirban.kundu.cu@gmail.com}
and
{\bf Shamayita Ray} $^{a,b,}$\footnote{shamayita.ray@gmail.com, sr643@cornell.edu}

\vspace{3mm}
 ${}^a$ \ \ {\em{Department of Physics, University of Calcutta, \\
92 Acharya Prafulla Chandra Road, Kolkata 700009, India
}}

${}^b$ {\em{Laboratory for Elementary Particle Physics, Cornell University, Ithaca, NY 14853, USA}}

\end{center}

\begin{abstract}

We investigate the parameter space of the Standard Model enhanced by a gauge singlet 
real scalar $S$. Taking into account all the theoretical and experimental constraints, 
we show the allowed parameter space for two different types of such singlet-enhanced 
Standard Model. For the first case, the scalar potential has an explicit $Z_2$-symmetry,
and may lead to a dark matter candidate under certain conditions. For the second case, the scalar 
potential does not respect any $Z_2$. This is again divided 
into two subcategories: one where the Standard Model vacuum is stable, and one where it 
is unstable and can decay into a deeper minimum. We show how the parameters 
in the scalar potential control the range of validity of all these models. 
Finally, we show the effect of one-loop correction on the positions and depths of the minima 
of the potential. 
\end{abstract}



\setcounter{footnote}{0}
\renewcommand*{\thefootnote}{\arabic{footnote}}

\section{Introduction}
\label{intro}

One of the minimalistic extensions of the Standard Model (SM) is that by
one or more gauge singlet real (or complex) scalar field(s). 
Motivations to introduce a singlet scalar to the SM are, amongst others: (i) to provide 
a viable cold dark matter (CDM) candidate through Higgs portal models \cite{portal-dm}, 
(ii) to make the electroweak phase transition a strong first-order one \cite{1107.5441,ew-others},
and (iii) to address the naturalness problem of the SM Higgs 
boson \cite{singlet-veltman}. 
Phenomenological aspects 
of such singlets have also been discussed in case of colliders 
\cite{barger-real,barger-cmplx,robens,dawson,collider-singlet}, 
and in the context of electroweak precision constraints \cite{precision}. Such a singlet 
with a mass around 750 GeV may also be responsible for the recently observed excess in the 
diphoton channel \cite{atlas-750,cms-750} if one adds vectorial fermions to the model. 

One often imposes a $Z_2$-symmetry on the scalar potential under which the SM particles 
are all even and the extra singlet(s) is(are) odd, which can make the 
lightest $Z_2$-odd particle a CDM candidate. 
In an analogous way to what happens in R-parity conserving supersymmetry and universal extra 
dimension models, a $Z_2$-symmetry on the scalar potential under which 
the singlet $S$ is odd, can in principle lead to the 
Higgs portal dark matter models where $S$ constitutes
the dark matter. A necessary condition for this is zero (or infinitesimally small)
vacuum expectation value (VEV) 
for $S$, so that it cannot mix with the SM doublet scalar $\Phi$. 
One must remember that this $Z_2$ is rather ad hoc, introduced just for 
the sake of having a CDM candidate. 

The nature of the tree-level scalar potential has also been discussed by several 
authors \cite{potential-tree}. The potential is more complicated than the 
SM one because of one extra field and the possibility that the CP-even neutral 
component of the SU(2) doublet $\Phi$, which will be denoted by $\phi$,  
and the gauge singlet scalar $S$ can both have nonzero VEVs. We denote these VEVs by $v$ 
and $v_s$ respectively. If there is only one minimum for $v_s$, one often uses 
the shift symmetry $S\to (S+\Delta)$ to 
ensure $v_s=0$, which in turn ensures a CDM candidate if the model is $Z_2$-symmetric.
However, if there are more than one minima for $v_s$, there is no particular advantage 
in using the shift symmetry, except ensuring that $v_s=0$ is an extremum.

In this paper, we investigate the nature of the tree-level potential of the 
SM extended by one real singlet scalar, which we call SM+S, 
with and without the $Z_2$-symmetry.
We find the allowed parameter space for three different SM+S models: 
(1) the Higgs portal model
{\em i.e.}, the $Z_2$-symmetric model where $S$ can be a CDM candidate, 
(2) the $Z_2$-symmetric model with no CDM candidate, and (3) the $Z_2$-asymmetric 
model. Obviously, the allowed parameter space has to be consistent with all theoretical and current 
experimental constraints. For the third case, $Z_2$ breaking is soft, coming 
from operators with mass dimension less than 4. Dimension-4 $Z_2$ breaking 
operators are forbidden from gauge and/or Lorentz symmetry.
As we consider only one real singlet scalar, all the couplings are real in these models.

Data from the Large Hadron Collider (LHC) 
essentially constrains the mixing between $S$ and $\phi$.
The mixing angle $\theta$ is constrained to be so small 
that all SM+S models also satisfy the constraints coming from the oblique parameters.
We will discuss it in detail later. The measurement of the $W$ boson mass also puts 
serious restrictions on the parameter space of SM+S \cite{1406.1043}.  

Apart from these experimental constraints, there are three other important constraints. 
Firstly, the potential has to be stable at all energy scales till one reaches the range 
of validity. This range is, in general, way below the Planck scale, apart from some exceptional 
choices of the parameters. Above this limit, either at least one of the couplings blow up, or the 
potential becomes unbounded from below along some direction in the field space. 
The second one is the existence of a minimum, either global or local, where $\bra\phi\ket=v=246$ GeV, 
which will be referred to as the electroweak (EW) vacuum. 
The third constraint comes from the stability of the EW vacuum; if there is another minimum deeper than 
the EW vacuum, the tunnelling lifetime should not be less than the age of
the universe.

We also study the effect of one-loop corrections to the potential. In general, the one-loop 
corrections are expected to be small compared to the tree-level potential, unless one 
looks along a flat direction. Even when both $S$ and $\Phi$ have non-zero mass terms, one can 
find a direction in the field space, at least for large values of the fields, following the 
prescriptions of Gildener and Weinberg \cite{gildener}. One should also choose the regularization
scale properly. This choice, in principle, is arbitrary if one uses renormalization group (RG) 
improved couplings. On the other hand, one can always tune the scale so that the one-loop corrected 
EW vacuum still has $\bra\phi\ket = 246$ GeV~\cite{indrani-4}. 
However, this makes the choice of the regularization scale dependent on the model parameters.

The paper is arranged as follows. In \autoref{model}, we give a brief outline of the singlet-enhanced 
SM, and discuss the constraints. \autoref{vs0} discusses the $Z_2$-symmetric model where the CDM is 
allowed, and \autoref{vsne0} is on models with a non-zero singlet-doublet mixing. While these are all 
tree-level results, we discuss the one-loop corrections to the potential in \autoref{oneloop}. In \autoref{summary},
we summerize and conclude.

\section{The real singlet scalar enhanced SM}
\label{model}

Let us consider the most general potential for SM+S, the single real-scalar extended SM:
\bea
V(\Phi,S) &=& -\mu^2\Phi^\dag\Phi - M^2 S^2 + \lambda\left(\Phi^\dag\Phi\right)^2 + a_1 \Phi^\dag\Phi S + 
a_2 \Phi^\dag\Phi S^2 
+ b_1 S + b_3 S^3 + b_4 S^4\,,
\label{pot}
\eea
where $\Phi$ is the SM doublet and $S$ is a gauge singlet scalar field.
We denote the CP-even 
neutral component of $\Phi$ by $\phi/\sqrt{2}$, and the VEVs are 
given as $\bra\phi\ket = v$, $\bra S \ket = v_s$. 
There are six new parameters in \autoref{pot} over and above those in the SM. Of them $a_2$, $M^2$, and $b_4$ 
respect the $Z_2$ symmetry of $S\to -S$, while
$a_1$, $b_1$ and $b_3$ break it softly. Note that $\mu^2,M^2 > 0$ stand for wrong-sign mass terms in the potential. 
All the couplings are, of course, real, because we have only one real singlet $S$ in this model.  

The stability conditions are obtained from the requirement that the potential should not become 
negative along any direction of the field space, which gives 
\be
\lambda > 0\,, \quad  b_4 > 0\,, \quad a_2 + 2\sqrt{\lambda b_4} > 0\,,
\label{stability}
\ee
along the directions $S=0$, $\Phi =0$, and $\sqrt{\lambda}\Phi^\dag\Phi 
= \sqrt{b_4}S^2$ directions respectively.

In case there is mixing between $\phi$ and $S$, 
the mass eigenstates $(h,s)$ are defined as
\be
h = \phi \cos\theta + S \sin\theta\,, \quad
s = -\phi\sin\theta + S \cos\theta\, ,
\label{mass-eigen}
\ee
where $\theta$ is the mixing angle. 

In terms of VEVs $v$ and $v_s$, the potential in \autoref{pot} becomes
\be
V(v,v_s) = -\frac12 \mu^2 v^2  - M^2 v_s^2 + \frac14\lambda v^4  + \frac12 a_1 v^2 v_s + 
\frac12 a_2 v^2 v_s^2 +  b_1 v_s + b_3 v_s^3 + b_4 v_s^4\,,
\label{pot-v-vs-1}
\ee
and the extremization conditions are
\bea
-\mu^2 v + \lambda v^3 + a_1 vv_s + a_2 vv_s^2 &=& 0\,, \label{pot-min1-v}  \\
\frac12 a_1 v^2 + a_2 v^2 v_s  + b_1 - 2M^2 v_s + 3b_3 v_s^2 + 4b_4 v_s^3 &=& 0 \, .
\label{pot-min1-vs}
\eea

One can always apply the 
shift symmetry $S\to (S+\Delta)$ to \autoref{pot}, where $\Delta$ is some constant, 
since this shift in $S$ does not change the physics. We can use this freedom\footnote{One can 
always remove the tadpole term if the potential is $Z_2$-symmetric. However, for a $Z_2$-asymmetric 
potential, there can be two minima with different values of $v$, and removal of the tadpole at one minimum does not 
ensure its removal at the other.} to remove one of the independent terms of the potential. Let us 
choose 
\be
b_1 = -\frac{1}{2} a_1 v^2 \, .
\label{b1}
\ee
Use of \autoref{b1} in \autoref{pot-v-vs-1} removes the linear terms in $S$
and simplifies the potential to
\be
V(v,v_s) = -\frac12 \mu^2 v^2  - M^2 v_s^2 + \frac14\lambda v^4  + 
\frac12 a_2 v^2 v_s^2 +  b_3 v_s^3 + b_4 v_s^4\, .
\label{pot-v-vs}
\ee
The extremization conditions now guarantee one extremum line along $v=0$ and 
another along $v_s=0$ (because of the shift symmetry):
\bea
v \left(-\mu^2  + \lambda v^2 + a_2 v_s^2\right)  &=& 0\,, \label{pot-min2-v}\\
v_s \left(a_2 v^2   - 2M^2  + 3b_3 v_s + 4b_4 v_s^2 \right) &=& 0\,.
\label{pot-min2-vs}
\eea
From Eqs.\ \eqref{pot-min2-v} and \eqref{pot-min2-vs} one finds that 
there are three extrema for $v_s$, and for each $v_s$ there are three extrema for $v$, 
depending on the existence of real 
solutions\footnote{For any given value of $v_s$, there is one extremum at 
$v=0$. If \autoref{pot-min2-vs} has three real 
solutions, and \autoref{pot-min2-v} has three real solutions for each of the solutions for $v_s$, 
there can be nine such extrema. Out of these nine extrema, three have $v=0$ and thus cannot be the 
EW vacuum. The two solutions for nonzero $v$ for a given solution of $v_s$ are symmetrically placed 
about $v=0$.}. For $v=0$, \autoref{pot-min2-vs} can be simplified to
\be
v_s \left(- 2M^2  + 3b_3 v_s + 4b_4 v_s^2 \right) = 0\,.
\label{pot-min3-vs-v0}
\ee
For the other two nonzero extrema in the $\Phi$-direction, one uses $v^2 = (\mu^2 - a_2 v_s^2)/
\lambda$ from \autoref{pot-min2-v} and get
\be
v_s \left( \left[4 b_4 - \frac{a_2^2}{\lambda}\right] v_s^2 + 3 b_3 v_s + \left[ \frac{a_2\mu^2}{\lambda} -2M^2\right]
\right) = 0\,.
\label{pot-min3-vs}
\ee
Given the parameters of the potential, Eq.\ (\ref{pot-min3-vs}) gives the condition for real 
non-zero solutions for $v_s$ along $v\ne0$. 
Note that for a $Z_2$-symmetric potential ($b_3=0$), the stability criteria ensure that 
$(a_2\mu^2/\lambda - 2M^2) < 0$. 

The condition that any extremum of $V(v,v_s)$ is a minimum, and not a maximum or saddle point, is 
\be
\frac{\partial^2 V}{\partial v^2} > 0\,, \quad 
\left[ \frac{\partial^2 V}{\partial v^2} \frac{\partial^2 V}{\partial v_s^2} - 
\left(\frac{\partial^2 V}{\partial v \partial v_s}\right)^2 \right] > 0\,.
\label{cond-minima}
\ee

To get the mixing between the CP-even component of the SM Higgs doublet $\phi$ 
and the real scalar $S$, we expand the potential in \autoref{pot} about the VEVs of the 
fields. The terms quadratic in fields are given by
\bea
V(\phi,S) &\supset& \phi^2\left[-\frac{\mu^2}{2} + \frac32 \lambda v^2 + \frac{a_1}{2} v_s + \frac{a_2}{2}v_s^2\right]
+ S^2 \left[ -M^2 + \frac{a_2}{2}v^2 + 3 b_3 v_s + 6 b_4 v_s^2\right] 
+ 2\phi S\left[\frac{a_1}{2} v + a_2 vv_s\right] \nonumber\\
&\equiv &\begin{pmatrix}\phi & S\end{pmatrix}
 {\cal M} \begin{pmatrix} \phi \cr S\end{pmatrix} \; ,
\eea
where
\bea
{\cal M} = 
\begin{pmatrix}
  \frac12\left( -{\mu^2} + 3 \lambda v^2 + {a_1} v_s + {a_2}v_s^2\right)  &   \frac{a_1 v}{2} + a_2 vv_s\cr
\frac{a_1 v}{2} + a_2 vv_s &  -M^2 + \frac{a_2v^2}{2} + 3 b_3 v_s + 6 b_4 v_s^2 \end{pmatrix} \,,
\label{massmat}
\eea
in its most general form. It should be noted that the mixing between $\phi$ and $S$ also depends on 
the parameter $a_1$, which does not appear in the minimization conditions in \autoref{pot-min2-v} and 
\eqref{pot-min2-vs}, because of the choice in \autoref{b1}.
If $M_1, M_2$ are the eigenvalues of ${\cal M}$, the masses of the physical states are given by 
\be
m_h = \sqrt{2 M_1} \; , \quad m_s = \sqrt{2 M_2}\,,
\label{mass}
\ee
and the mixing angle $\theta$ is the angle which parametrize the $2\times2$ 
rotation matrix that diagonalises ${\cal M}$.

\subsection{Zero VEV for at least one field}

One instructive case is when one of the minima has either $v=0$ or $v_s=0$ or both.
 
 \subsubsection{Minimum at $v=v_s=0$}
 \label{v0-vs0}
 
The point $v=v_s=0$ is an extremum where
\bea
\frac{\partial^2V}{\partial v^2} = -\mu^2 \; , \quad \frac{\partial^2V}{\partial v_s^2} = -2 M^2 \; , 
\quad \frac{\partial^2V}{\partial v \partial v_s} = 0 \; , 
\label{extreme0}
\eea
and will be a minimum if $\mu^2, M^2 < 0$ i.e. the mass terms are right-sign. 

From \autoref{pot-min2-v} one can see for this case that $v$ will also have two non-zero real roots 
if $(a_2 v_s^2 - \mu^2) < 0$, which requires a large negative $a_2$, resulting in a large 
singlet-doublet mixing.

\subsubsection{Minimum at $v=0$, $v_s\ne0$}

If there is a minimum for $v=0$, $\partial^2 V/\partial v \partial v_s = 2a_2 v v_s
=0$, and the condition for minimum translates to
\be
\frac{\partial^2 V}{\partial v^2} > 0 \ \ \Rightarrow \ \   a_2 v_s^2 -\mu^2 > 0 \; ,\ \ \ 
\frac{\partial^2 V}{\partial v_s^2} > 0 \ \ \Rightarrow \ \ 8 b_4 v_s^2 + 3 b_3 v_s > 0\;.
\ee
The first condition shows, from \autoref{pot-min2-v}, that there is only one real solution for
$v$ at $v=0$. 
As $b_4 > 0$ from stability criterion, the second condition yields interesting bounds. 
If $b_3 < 0$ and $v_s < 0$ or $b_3 >0$ and $v_s>0$, 
${\partial^2 V}/{\partial v_s^2}$ is definitely positive. On the other hand, 
if $v_s > 0$ but $b_3 < 0$, there is a lower bound on $v_s$,
\be
v_s > \frac{3 |b_3|}{8 b_4}\,.
\ee
Similarly, if $b_3 > 0$ and $v_s < 0$, there is an upper bound on $v_s$:
\be
v_s < - \frac{3 b_3}{8 b_4}\,.
\ee

\subsubsection{Minimum at $v\ne0$, $v_s=0$}

For $v\not=0$ and $v_s=0$, the minima along $v$ are symmetric, at 
$v=\pm \sqrt{\mu^2/\lambda}$. This means $\mu^2 > 0$ for $v$ to be real 
and hence $\partial^2V/\partial v^2 = 2 \mu^2 > 0$. 
So the condition for minimum at $v\ne0$, $v_s=0$ reduces to $(a_2 v^2 - 2M^2) > 0$.

{\subsection{Minimum at $v\not= 0$, $v_s \not=0$}
\label{v-vs-nonzero}}

The constraints for $v\ne0$, $v_s\ne0$ are obtained using the concavity condition in 
\autoref{cond-minima}, along with 
the expression for the second derivatives, to be
\bea
\frac{\partial^2V}{\partial v^2} = 2\left(\mu^2 - a_2 v_s^2\right) = 2 \lambda v^2 \; , \quad 
\frac{\partial^2V}{\partial v_s^2} = 3 b_3 v_s + 8 b_4 v_s^2 \; , 
\quad \frac{\partial^2V}{\partial v \partial v_s} = 2 a_2 v v_s \; .
\eea
For the $Z_2$-symmetric potential, $b_3=0$, and hence the condition for a minimum at 
$v\ne0$,$v_s\not= 0$ simplifies to  
$4 \lambda b_4 > a_2^2$, which is nothing but one of the stability criteria for the 
potential as shown in \autoref{stability}.

Further simplifications occur if we have only one minimum of the potential. 
In this case, it has to be at $|v|=246$ GeV, and hence solutions 
for $v_s\ne 0$ can be obtained in a straightforward way from \autoref{pot-min2-vs}:
\be
v_s = \frac{-3b_3 \pm \sqrt{9 b_3^2 - 16b_4(a_2 v^2 - 2M^2)}}{8b_4}\,.
\label{vs-sol2}
\ee
Thus, for the $Z_2$-symmetric case, 
the condition for a minimum at $v_s\not= 0$ is 
$a_2 v^2 - 2M^2 < 0$, as $b_4 > 0$ from stability criterion. 
For nonzero $b_3$, the condition is
\be
a_2 v^2 - 2 M^2 < 9 b_3^2 / 16 b_4\,.
\ee

One can easily have more than one minima with $v\not=0$ and $v_s\not=0$. However, if one minimum is at the 
origin, the second minimum at nonzero $v$ and $v_s$ requires $a_2$ to be large and negative, 
as discussed in Section~\ref{v0-vs0}. Such large values of $a_2$ are under severe kosh from the LHC data. 
A detailed discussion on the nature of the scalar potential related to the electroweak phase transition can 
be found in Ref.\ \cite{1107.5441}.


\subsection{LHC constraints}

If there is no mixing between $\phi$ and $S$, there are no constraints on $S$ coming from 
the LHC data, except that it cannot be so light ($<m_h/2$) 
that $h \to SS$ is allowed and the branching ratio
is more than $34\%$ at 95\% CL \cite{atlas-cms-combo}. Similarly, 
there is no constraint from electroweak precision observables as $S$ does not have any 
gauge coupling.

If there is a mixing between $\phi$ and $S$ parametrized 
by an angle $\theta$, the 
number of events, which is just the production cross-section times the branching ratio, 
goes down by $\cos^2\theta$. 
Denoting the production cross-section times the decay width scaled to that in the SM by 
$\mu$, the ATLAS and CMS combined result shows 
\be
\mu = 1.09^{+0.11}_{-0.10}\,,
\ee
from which we can put a limit of $\theta \leq 0.1$ at $1\sigma$, which we will use for our 
subsequent discussion. This helps us to avoid both 
LHC data and precision constraints at one stroke.

\subsection{Oblique parameters}

Only the $T$ parameter may be significant in the small mixing case. In this model, 
the $T$ parameter is given by \cite{barger-real}
\bea
T^{\rm{SM+S}} &=& -\left(\frac{3}{16\pi s_W^2}\right) 
\left\{ \cos^2\theta\left[ \frac{1}{c_W^2}\left(\frac{m_1^2}{m_1^2-m_Z^2}\right) \, 
\ln\frac{m_1^2}{m_Z^2} - \left(\frac{m_1^2}{m_1^2-m_W^2}\right) \, 
\ln\frac{m_1^2}{m_W^2}\right] \right. \nonumber\\
&& \left. + \sin^2\theta \left[ \frac{1}{c_W^2}\left(\frac{m_2^2}{m_2^2-m_Z^2}\right) \, 
\ln\frac{m_2^2}{m_Z^2} - \left(\frac{m_2^2}{m_2^2-m_W^2}\right) \, 
\ln\frac{m_2^2}{m_W^2}\right]\right\}\,,
\eea
where $m_1 (\approx 125$ GeV) and $m_2$ are the two mass eigenstates, and $\theta$ is the mixing angle. 
The SM expression for $T$ can be found by putting $\theta=0$ and $m_1=m_h$.
The quantity constrained by the electroweak fit, $\Delta T$, is given by \cite{pdg2014-ewreview}
\be
\Delta T = T^{\rm{SM+S}} - T^{\rm{SM}} = 0.01 \pm 0.12\,.
\ee
$T$ is related with the $\rho$-parameter by $\rho-1=\alpha T$. For small mixing ($\theta \leq 0.1$), 
the constraints coming from the oblique parameters are not significant.

\subsection{Renormalization Group equations}

We would also like to see how the couplings evolve with energy. The one-loop $\beta$-functions are 
\cite{Chakraborty:2012rb}
\bea
16\pi^2 \beta_\lambda &=& 12\lambda^2 + 6g_t^2\lambda + a_2^2 -\frac32\lambda(g_1^2+3g_2^2) 
- 3 g_t^4 + \frac{3}{16}(g_1^4 + 2 g_1^2 g_2^2 + 3 g_2^4)\,,\nonumber\\
16\pi^2 \beta_{b_4} &=& 36 b_4^2 + a_2^2 
\,,\nonumber\\
16\pi^2 \beta_{a_2} &=& \left[ 6\lambda + 12 b_4 + 4a_2 + 6g_t^2  -\frac32 g_1^2 -\frac92 g_2^2\right]\; a_2
\,,\nonumber\\
16\pi^2 \beta_{g_t} &=& \left[ \frac94 g_t^2 - \frac{17}{24}g_1^2 - \frac98 g_2^2 - 4g_3^2\right]\; g_t\,
\label{all-rge}
\eea
where $\beta_h\equiv dh/dt$, and $t \equiv \ln(Q^2/\mu^2)$. The $\beta$-functions for all gauge couplings 
are identical to that of the SM. 
For simplicity, we have put all the SM Yukawa 
couplings equal to zero except for that of the top quark. This hardly changes our conclusions. 

For the $Z_2$-asymmetric case, the trilinear couplings $a_1$ and $b_3$ also evolve:
\bea
16\pi^2 \beta_{a_1} &=& a_1\left(9\lambda + 4 a_2\right) + 6 a_2 b_3\,,\nonumber\\
16\pi^2 \beta_{b_3} &=& 2 a_1 a_2 + 36 b_3 b_4\,.
\label{z2-odd-rge}
\eea

Note that $\beta_{b_4}$ is always positive and hence can only increase, starting from a positive value. 
$\beta_\lambda$ also gets a positive contribution on top of the SM ones. These make the couplings blow 
up at a much lower scale than the Planck scale ($\sim 10^{19}$ GeV), unless one starts with very small 
values of $b_4$. Similarly, $\beta_{a_2}$ is proportional to $a_2$ itself and can lead to a blow-up 
for large $a_2$. 

For our analysis, we have taken the initial values of the couplings at the electroweak scale in such a way that 
the Higgs boson mass is correctly reproduced as $m_h\in [124:126]$ GeV. The threshold effects are taken at the 
singlet mass scale, however it has been seen that the final results are not very sensitive on the exact choice of 
this scale, and moreover, uncertainties coming from 
possible higher-loop contributions are larger compared to the uncertainties coming from the threshold 
corrections. The constraints on the parameter space are all obtained with the one-loop improved values of the 
couplings. 

\subsection{The unstable vacuum case}

If there are two minima of the potential and the EW vacuum is shallower, the universe can tunnel down
to the deeper vacuum. In such cases, the parameters must be chosen such that the lifetime of the shallower vacuum 
should be at least as large as the lifetime of the universe, which is about 13.7 billion years. 

\begin{figure}[t!]
\begin{center} 
\includegraphics[width=7cm,height=5.8cm]{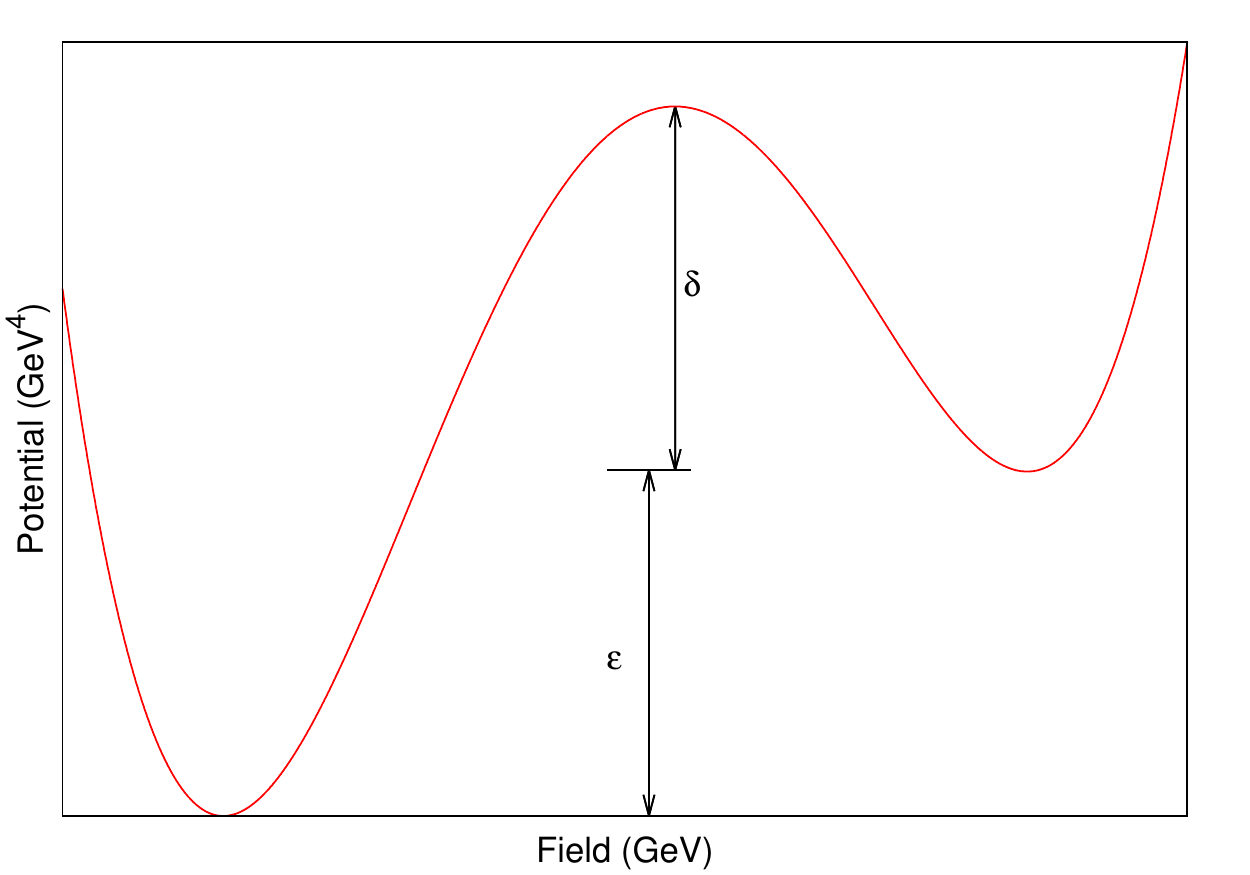}
\includegraphics[width=7cm,height=5.6cm]{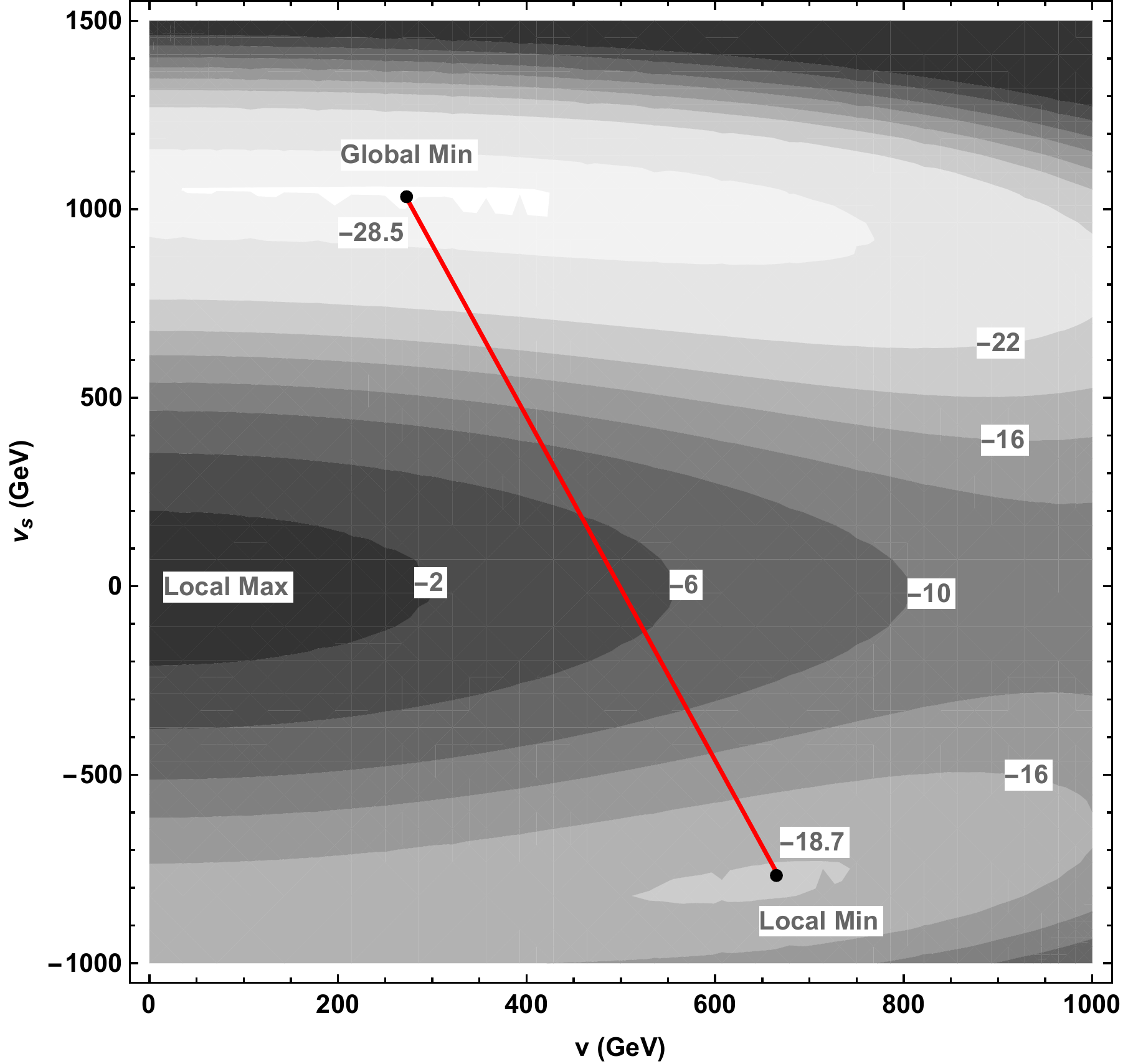}
\end{center} 
\caption{\small (a) Left: Schematic diagram of an asymmetric double-well potential with a single field. 
Parameters $\epsilon$ and $\delta$ control the tunnelling lifetime. If there are more than 
one fields, one gets a multi-dimensional contour. (b) Right: Example of a two-dimensional contour for 
SM+S. The line shows the shortest path joining the two minima along which the tunnelling probability 
should be calculated. Notice that the path does not pass through the local maximum. Contour values denote the 
potential at that point in the unit $10^{10}$ GeV$^4$.}
\label{fig:metastable}
\end{figure}

To calculate the lifetime of the metastable state \cite{coleman}, let us assume that the potential 
at the EW vacuum is zero (this can always be achieved by a constant shift), the true vacuum has a depth 
$-\epsilon$ ($\epsilon > 0$), and the height of the barrier with respect to the SM vacuum is 
$\delta$, as shown in \autoref{fig:metastable}. The decay width density of the universe, $\Gamma/V$, is given by 
\be
\frac{\Gamma}{V} = A \exp(-B)\,,
\ee
where $A$ is a small pre-factor, and $B$ is estimated in the thin-wall approximation as 
\cite{barroso}
\be
B = \frac{2^{11}\pi^2}{3\lambda} \left( \frac{\delta}{\epsilon}\right)^3\,.
\label{B-def}
\ee
This expression is true for a single field whose self-quartic coupling is given by $\lambda$. 
With more than one fields, there are several dimensionless couplings and the tunnelling path may not 
even go through under the hill. However, for SM+S, what we have done numerically is to find the 
two minima and then calculate the tunnelling along the straight line joining them, as shown in 
\autoref{fig:metastable}. This is effectively a single-field approximation, where the field is a 
combination of $h$ and $S$. The height $\delta$ is taken to be the maximum height above zero-level 
{\em along this path} and not the local maximum of the field space. 

The ratio $\delta/\epsilon$ should be greater than $0.1$ to make the shallower vacuum stable 
with respect to the lifetime of the universe. Keeping in mind of the necessary simplifications, 
we have chosen only those models for which this ratio is more than unity, and therefore the stability is 
assured.

\subsection{One-loop corrections to the potential}

 The one-loop effective potential in the SM is given by

 \be
 V_1(\phi_c) = \frac{1}{64\pi^2} \sum_i n_i m_i(\phi_c)^4 \left( \ln\frac{m_i(\phi_c)^2}{Q^2} - C_i\right)\,,
 \ee
 where $i$ runs over $h,G,W,Z$ and $t$, and
 \be
 n_i = 1,3,6,3,-12\,,  \ \ 
 C_i = \frac32,\frac32,\frac56,\frac56,\frac32\,,
 \ee
 for $i=h,G,W,Z$ and $t$ respectively. The masses are field-dependent and depend on the classical minimum $\phi_c$.
 $Q$ is the arbitrary regularization scale.  
 
 To be precise, the radiative corrections, being suppressed by the loop factor, are significant only along a 
 flat or near-flat direction in the field space. If all the dimensionful parameters are zero, the theory becomes 
 scale invariant. The minimization conditions in this case, namely, $v \left(\lambda v^2 + a_2 v_s^2 \right)= 0$ and 
 $v_s \left( a_2 v^2 + 4 b_4 v_s^2\right) = 0$, yield $4\lambda b_4 = a_2^2$ as the consistency condition if
 neither $v$ nor $v_s$ vanishes. The last condition makes the determinant of the mass matrix equal to zero, 
 ensuring a massless mode and hence a flat direction in the field space. If dimensionful couplings are 
 present, there is in general no flat direction in the $\phi$-$S$ plane, and one expects the radiative corrections
 to the potential to have a small effect. Note that 
 $4 \lambda b_4 = a_2^2$ is the limiting case
 of the stability condition. 
  However, If there is a strong hierarchy between 
 the two VEVs, the direction along the smaller-VEV field is almost flat. The choice of the regularization scale may change 
 $v_s$ significantly, and if there is a significant singlet-doublet mixing, both the physical scalar masses 
 may get affected. We will see the numerical estimates later. 
 
This also shows that putting $a_1 = - 2 a_2 v_s$ in \autoref{massmat} does not lead to a CDM candidate, 
because such a fine-tuned 
relationship is not stable under radiative corrections. 

With two scalar fields, the form of the one-loop corrected potential is
\be
V(\phi_c,\eta_c) = V_0(\phi_c,\eta_c) + V_1(\phi_c,\eta_c)\,,
\ee
where $\phi_c$ and $\eta_c$ are the classical minima along $\phi$ and $S$ respectively, and
$V_0$ is the tree-level potential written in terms of the classical minima. $V_1$ is the 
one-loop correction, given by
\be
V_1(\phi_c,\eta_c) = V_1(\phi_c) + \frac{1}{64\pi^2} m_S^4(\phi_c,\eta_c) \left[\ln \frac{m_S^2(\phi_c,\eta_c)}{Q^2} 
-\frac32\right]\,.
\ee
One should note that in case of nonzero mixing, the field-dependent mass of $h$ in $V_1(\phi_c)$ 
is a function of $\eta_c$ too. Thus, $V_1$ 
includes all the SM fields and the singlet. 

The regularization scale $Q$ is arbitrary, but one can choose it in such a way that the one-loop corrected minimum 
 for $\Phi$ remains unchanged, in other words, 
 $\phi_c$ falls at $v$. This keeps all the SM fermion and gauge boson masses invariant, and also keeps the Goldstone
 bosons massless; thus, we do not need to consider the Goldstone boson contributions for the effective potential. 
 A detailed discussion of the procedure is given in Ref.\ \cite{indrani-4} for the two-Higgs doublet model potential.

\section{${Z_2}$-symmetric potential with ${v_s=0}$}
\label{vs0}

Let us first go through the well-studied case of the $Z_2$-symmetric potential ($a_1=b_1=b_3=0$ in 
\autoref{pot-v-vs-1}), for the sake of completeness of this study. 
One can always get a minimum at $v_s=0$ by applying the shift symmetry 
$S\to S+\Delta$. This minimum should better be the only one, or at least the global one (a local one 
with lifetime more than the age of the universe will also do) if we want to have a 
CDM candidate in $S$, with mass of  $\sqrt{a_2 v^2 - 2M^2}$. The points with $M^2 > 0$ need a very large 
and negative $a_2$ and are ruled out by the CDM spin-independent scattering cross-section limits. 
Thus, all such Higgs portal dark matter models must have a right-sign mass term ($M^2 < 0$) for $S$. This is 
true even for the narrow region of Higgs resonance, at about $m_S\approx m_h/2$.

\begin{figure}[t!]
\begin{center}
\includegraphics[width= 8.2cm]{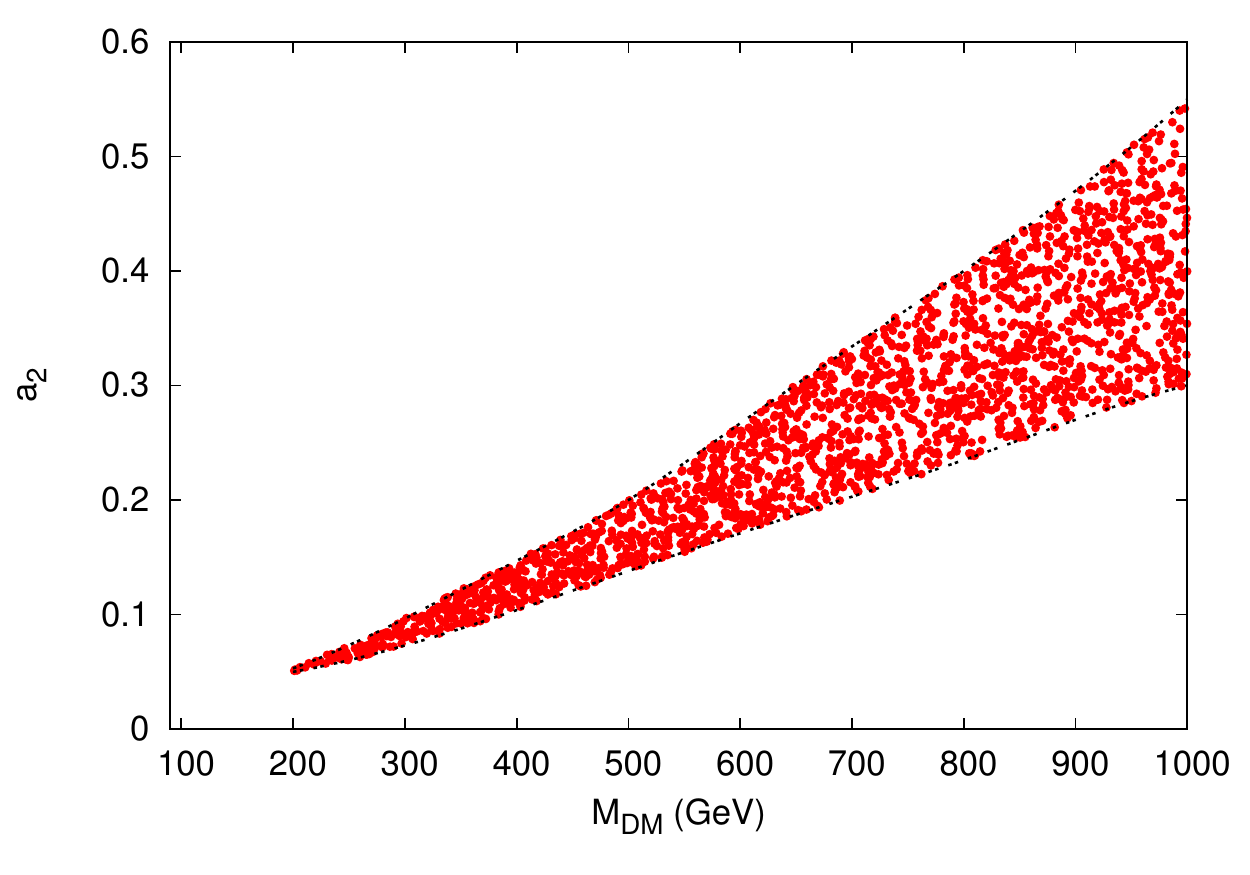}
\includegraphics[width= 8.2cm]{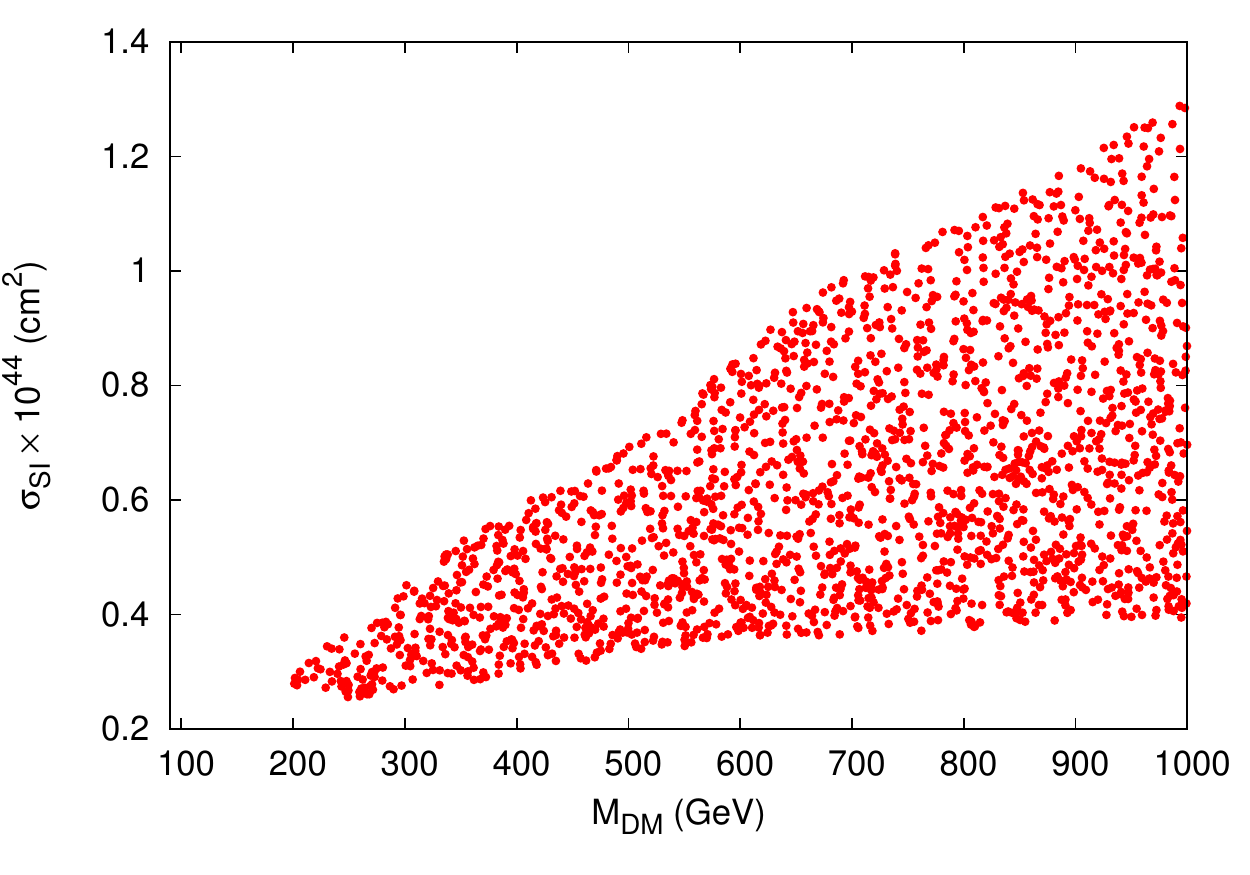}
\end{center}
\caption{\small Allowed values of the parameter $a_2$ and the spin-independent cross-section of the DM 
in the $Z_2$-symmetric case with the minimum at $v_s=0$.
(a) Left: Allowed region for $a_2$. Regions outside the wedge are ruled out from 
direct detection experiments and overclosure ($\Omega > \Omega_{CDM}$).
(b) Right: The spin-independent cross-section of the DM as a function of the DM mass. 
$a_2$ has been varied over the allowed range.}
\label{fig:dmallowed1}

\end{figure}

\begin{figure}[htbp]
\begin{center}
\includegraphics[width=8.5cm]{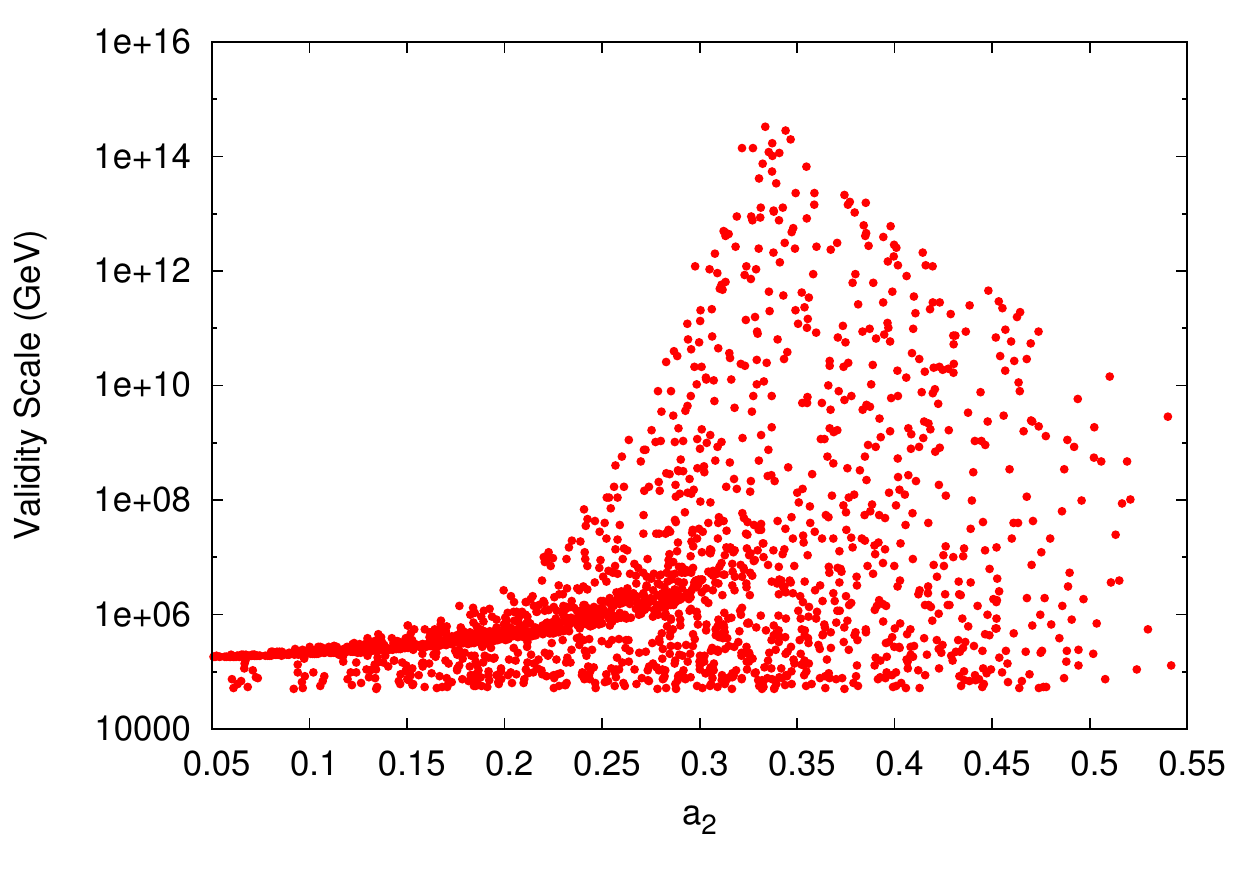}
\includegraphics[width=8.5cm]{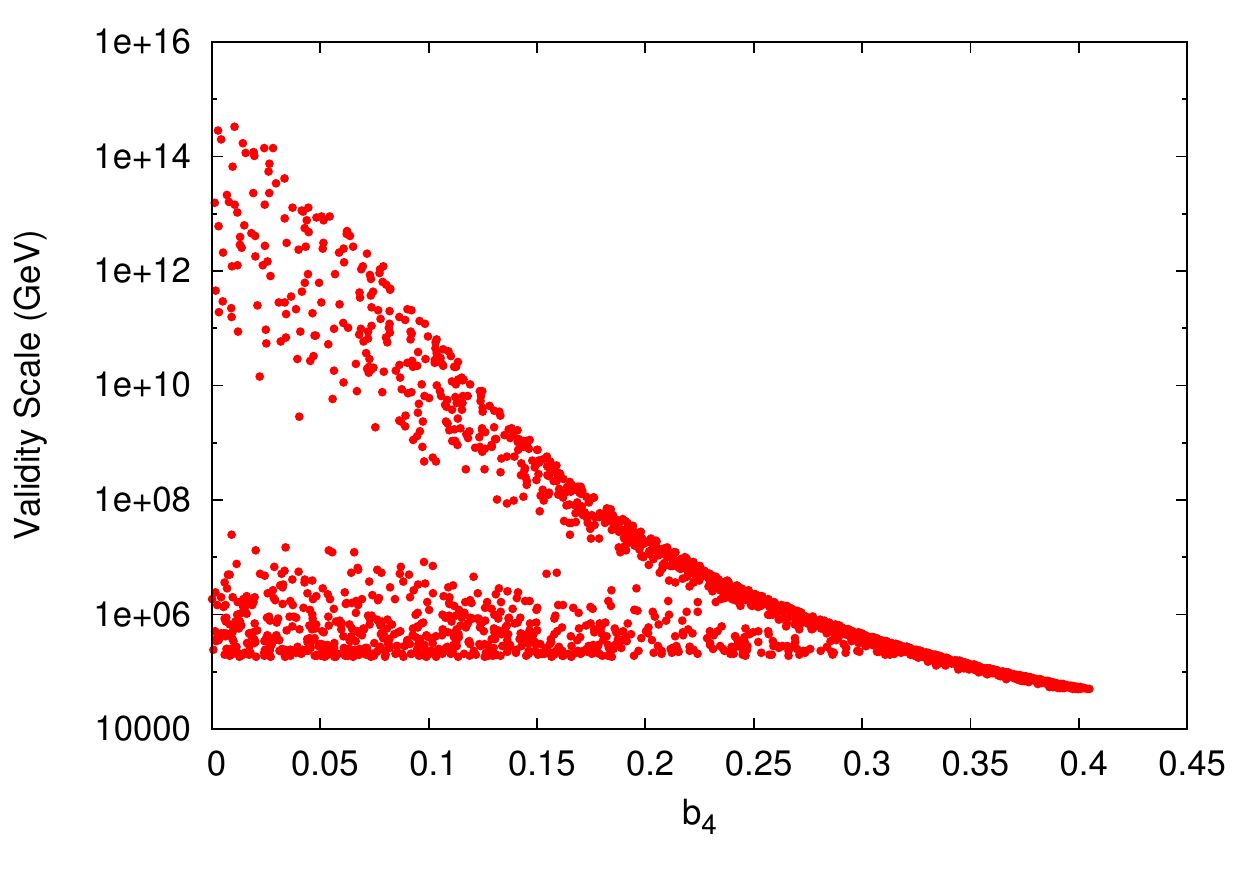}
\end{center}
\caption{\small The range of validity as a function of $a_2$ (left) and $b_4$ (right), in the $Z_2$-symmetric case with the minimum at $v_s=0$.}
\label{fig:DMvalidity}
\end{figure}

The only way for $S$ to interact with the SM sector is through the term $a_2 S^2 \Phi^\dag\Phi$ in \autoref{pot}, 
since $a_1 = b_1 = b_3 = 0$ in this case. The spin-independent 
CDM-nucleon scattering cross-section is given, in this scenario, by \cite{feng,duerr}
\be
\sigma = \frac{a_2^2 f^2 m_N^4}{\pi m_S^2 m_h^4}\,,
\ee
where $m_N, m_S$ and $m_h$ are the masses of the nucleon, $S$, and Higgs respectively. The matrix element 
for scattering is given by $f$, whose value is approximately $0.3$. 
If $a_2$ is very small, the CDM detection cross-section becomes small, but the annihilation rate goes down too, 
leaving more dark matter in the universe than is allowed, and thus leading to overclosure. 
If $a_2$ is large, the scattering cross-section of CDM with nucleons is also large and hence will be 
severely constrained by direct detection experiments, in particular LUX, which gives the best limits now 
\cite{dm-limits,lux}. Thus, apart from the narrow Higgs resonance region, 
only a small wedge for the dark matter mass $M_{\rm DM} > 200$ GeV is still allowed, 
and we focus only on those models that provide $M_{\rm DM}$ in this range.

In \autoref{fig:dmallowed1}, we show the allowed regions as a function of $a_2$, every dot corresponds to 
a particular choice of parameters. For all these models $M^2 < 0$, which, by our definition of the 
potential, means a right-sign mass term for the singlet and no symmetry breaking in the $S$-direction. 
Technically, $M^2 > 0$ can also lead to a local minimum at $\bra S \ket = 0$, 
but if we take $a_2$ to be in the perturbative region, such models lead to low $M_{\rm DM}$ and 
are hence ruled out by the direct detection data.

Depending on the values of $a_2$ and $b_4$, one can also check how far in the energy scale 
the singlet DM model remains valid. For this, we use the one-loop renormalization group (RG) 
equations\footnote{This gives a pretty good estimate, although two-loop results are available.} 
and see where the couplings become non-perturbative and ultimately hit the Landau pole, or the 
potential becomes unstable. Our results are shown in \autoref{fig:DMvalidity} for the two 
relevant parameters $a_2$ and $b_4$. Note that there is an upper limit on $b_4 \sim 0.4$ 
above which the model ceases to be valid even before 50 TeV. While there is no such limit 
for $a_2$, low-$a_2$ models have a smaller range of validity compared to medium-$a_2$ 
($\sim 0.3$) models, where the validity can be as high as $10^{15}$ GeV. 

While the allowed region for each model depends on the exact values of the parameters chosen, 
some intuitive insights can be put forward. As there is no mixing, $\lambda$ must start from its SM value
$\sim 0.13$. The only modification to its $\beta$-function comes from the $a_2^2$ term, so the 
range of validity increases with increasing $a_2$, provided $b_4$ is sufficiently small to start with 
and does not hit its Landau pole earlier ($b_4$ starts from a positive value, and always increases). 
So, for the low-$a_2$ regions, it is the vacuum stability that mostly controls the allowed range. 
After $a_2$ passes a certain value, and/or $b_4$ becomes large,
the range is controlled by the blowing up of one or more of the couplings. 
As we have just shown, $a_2 < 0$ is already ruled out for this model.

A similar study on the parameter space of Higgs portal dark matter models was performed recently 
in Ref.\ \cite{1509.01765}. There is always a chance that with improved measurements, the wedge region 
may go away. In the Higgs resonance region, the allowed values of $a_2$ are much smaller, and from 
\autoref{fig:DMvalidity}, we see that such models cease to be valid at about $10^6$ GeV. 

This model, with addition of vectorlike fermions, may explain the recently observed resonance 
at 750 GeV. However, introduction of such fermions spoils the possibility of a Higgs portal 
dark matter, as the scalar decays through the fermion loops. On the other hand, there is a
new constraint on the singlet mass, which narrows down the parameter space even further. The 
renormalization group equations also change, with new Yukawa couplings introduced, and may affect 
the stability of the potential. We will not discuss this extension any further here.


\section{Singlet-doublet mixing with $v_s\not=0$}
\label{vsne0}

\subsection{$Z_2$-symmetric case}
\label{z2symm_NoDM}

\begin{figure}[t!]
\begin{center} 
\includegraphics[width=8.5cm]{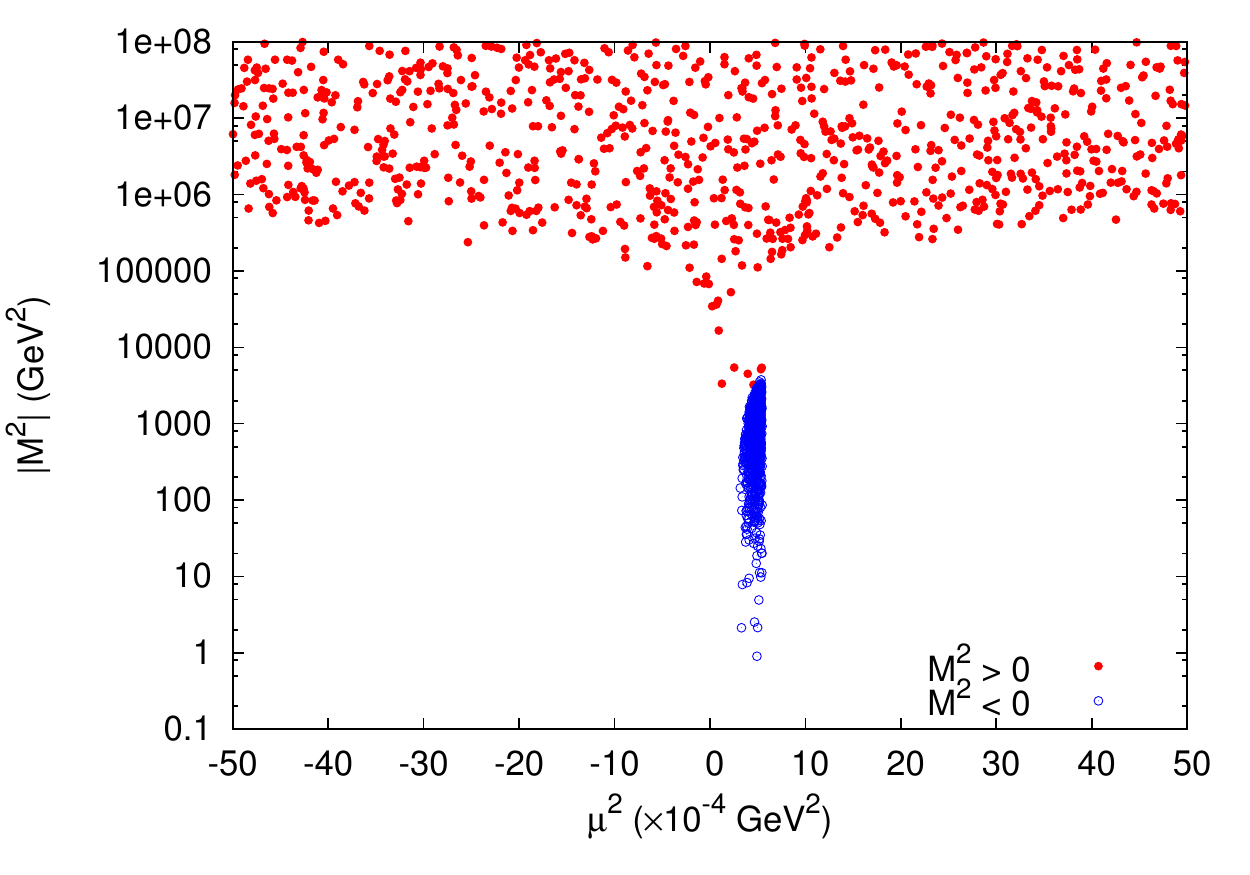}
\includegraphics[width=8.5cm]{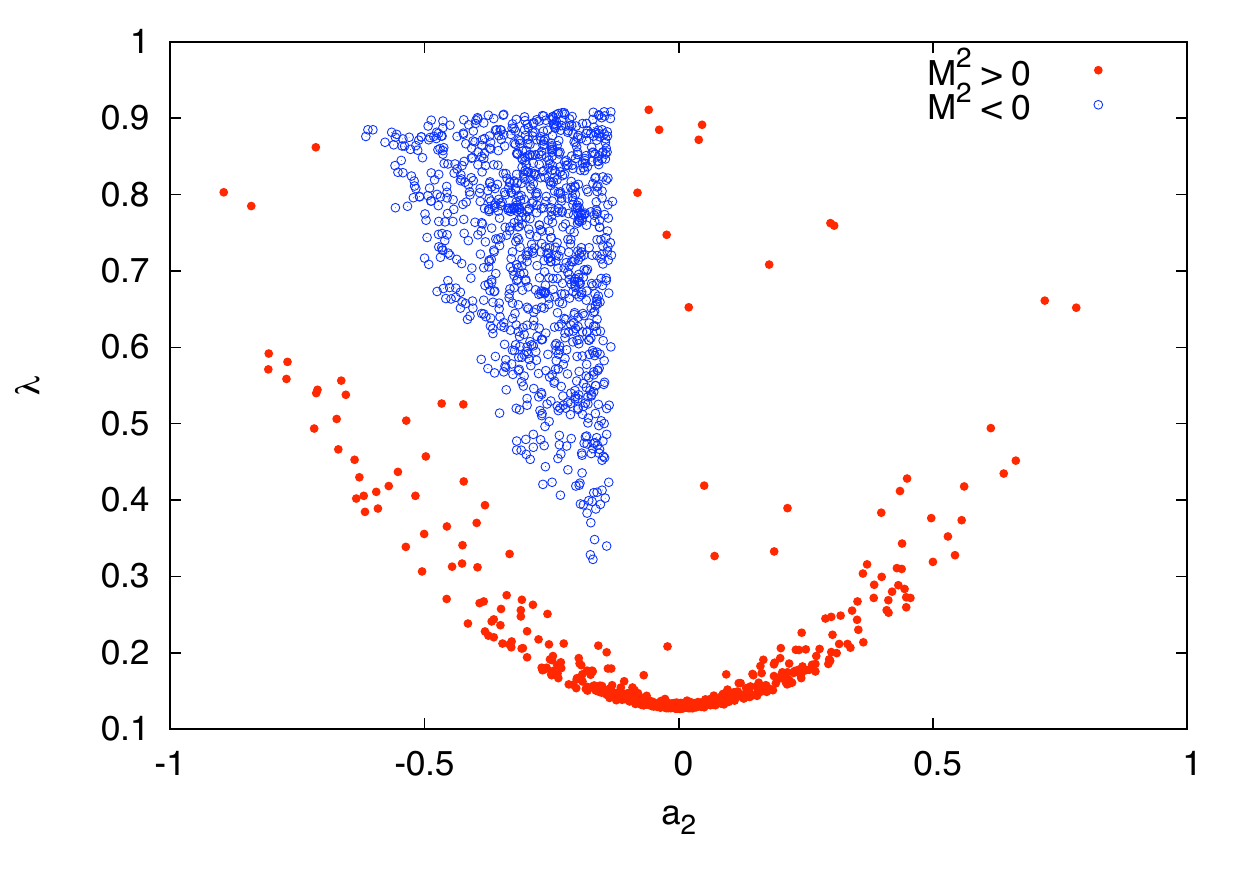}
\end{center} 
\caption{\small The allowed parameter space for $Z_2$-symmetric case with $v_s\not=0$.}
\label{fig:Z2noDM-aps}
\end{figure}
\begin{figure}[t!]
\begin{center}
\includegraphics[width=8.5cm]{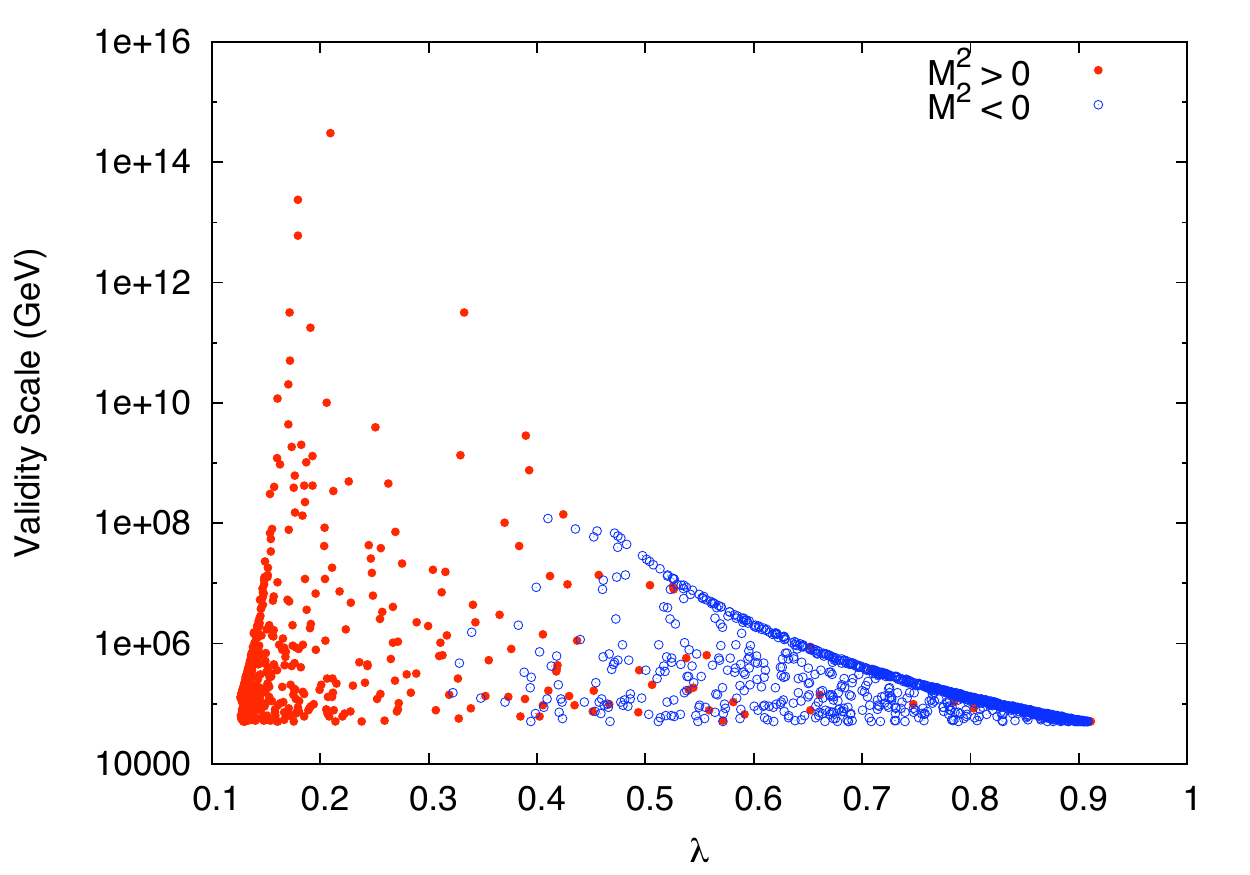}
\includegraphics[width=8.5cm]{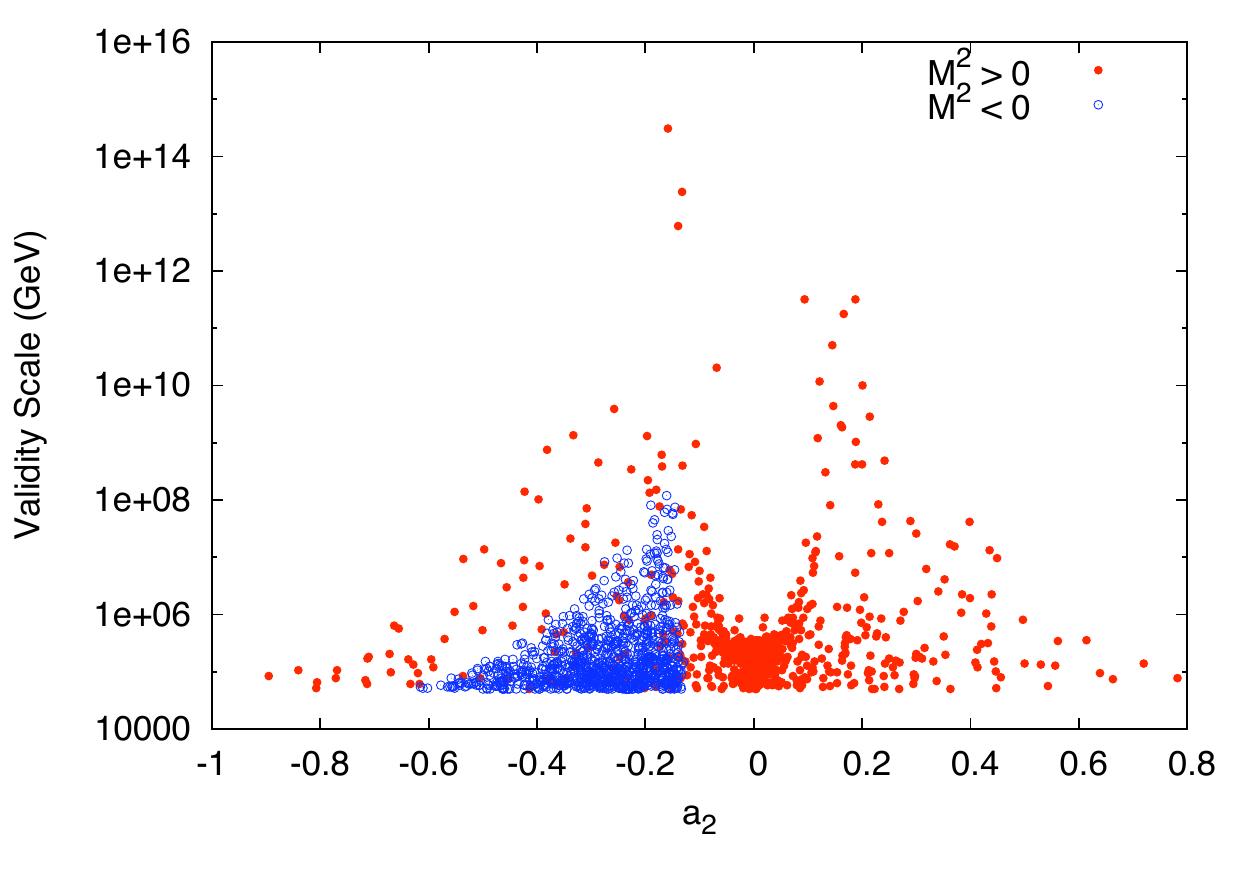}
\end{center} 
\caption{\small Range of validity for different values of $a_2$ and $\lambda$ in the $Z_2$-symmetric 
SM+S models with $v_s\not= 0$. The plot for $b_4$ is similar to that for $\lambda$. Large values of either 
$b_4$ or $\lambda$ limit the range of validity of the models, because of the nature of the 
RG equations. $a_2$ plays a subdominant role here.}
\label{fig:Z2noDM-val}
\end{figure}

If the singlet field $S$ develops a nonzero VEV, {\em i.e.} $v_s\ne0$, the physical fields $h$ and $s$ 
become orthogonal combinations of $\phi$ and $S$, with the mixing angle constrained by the 
LHC data to be $\theta < 0.1$. Hence there is no CDM candidate in this model \footnote{Such a model 
suffers from the usual domain wall problem. However, existence of multiple vacuum states in the universe 
is not yet ruled out, and in fact is a distinct possibility in several string theory motivated scenarios.}. 
The allowed 
parameter space is shown 
in \autoref{fig:Z2noDM-aps} for two distinct cases: $M^2 < 0$ (right-sign mass term for 
the singlet) and $M^2 > 0$ (wrong-sign mass term). We note the following characteristics:
\begin{itemize}

\item For $M^2 < 0$, only negative values of $a_2$ are 
allowed. This follows from the condition $(a_2 v^2 - 2M^2) < 0$. 
Both positive and negative values of $a_2$ are allowed for $M^2 > 0$.  

\item There is a correlation between $a_2$ and $\lambda$ for $M^2 < 0$. 
This follows from the $a_2^2/\lambda$ dependence of $v_s^2$ in \autoref{pot-min3-vs}. 

\item Only small positive values of $\mu^2$ are allowed for $M^2 < 0$. This can be understood from
the constraint $(a_2\mu^2/\lambda - 2M^2 )< 0$ coming from the stability criteria, as mentioned
in Sec.~\ref{model}. It can be shown easily that for $M^2 < 0$, $a_2$ and $\mu^2$ have to be of 
opposite signs to have real $v_s$, and the only possibility to have a real $v$ as well is to have $\mu^2 >0$ 
and $a_2<0$. The magnitudes of $a_2$, $\mu^2$ and $M^2$ are, however, restricted by the 
Higgs mass and the mixing angle $\theta$. For 
$M^2 < 0$, there is no such constraint and $|\mu^2|$ can be large.  

\end{itemize}
The range of validity for the models is shown in \autoref{fig:Z2noDM-val}. One can see that
large values of $a_2$ or $\lambda$ necessarily mean a smaller range of validity, which follows 
from the nature of the RG equations. The coupling $b_4$ always increases, so one has to start with 
a sufficiently small value of $b_4$ not to hit the Landau pole. The other two couplings, $\lambda$ and 
$a_2$, can be controlled by the negative contributions coming from Yukawa or gauge couplings if they 
are sufficiently small to start with. Too small a value means instability setting in at a low scale, 
and too large a value means a quick blowing up of the couplings. Thus, an intermediate range,
$a_2 \sim \lambda \sim 0.2$ is the region for maximum validity. 
We have explicitly checked that the mixing angle is always well within the LHC limit.

One-loop corrections do not change the nature of the potential qualitatively, except changing the depth 
of the potential. However, the composition of the CP-even neutral scalars change with the one-loop 
corrections, because the ratio of $\phi_c$ and $\eta_c$, and hence the mixing angle $\theta$ of the 
mass matrix changes from its tree-level value.


\subsection{$Z_2$-asymmetric case}
\label{z2asymm}

\begin{figure}[t!]
\centering
\includegraphics[width=8.5cm]{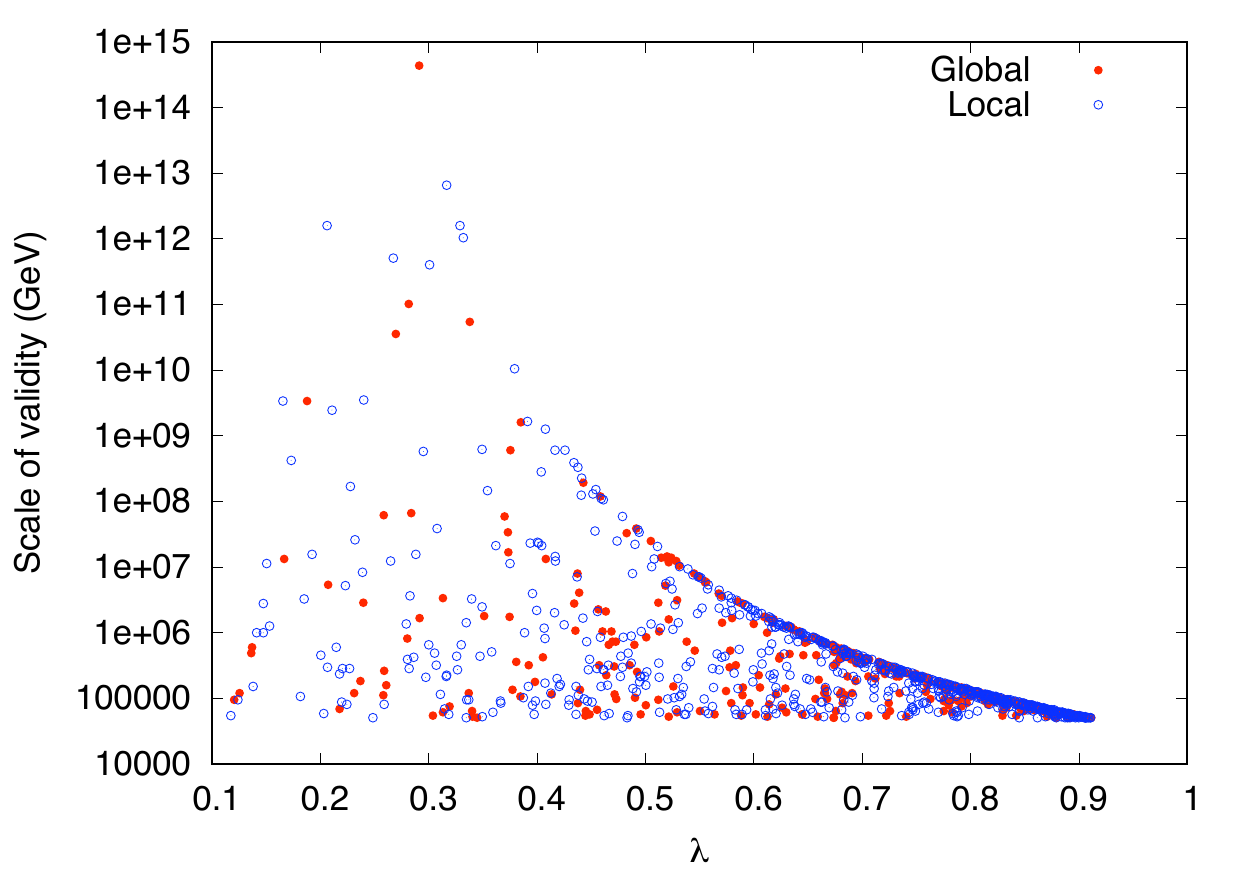}
\includegraphics[width=8.5cm]{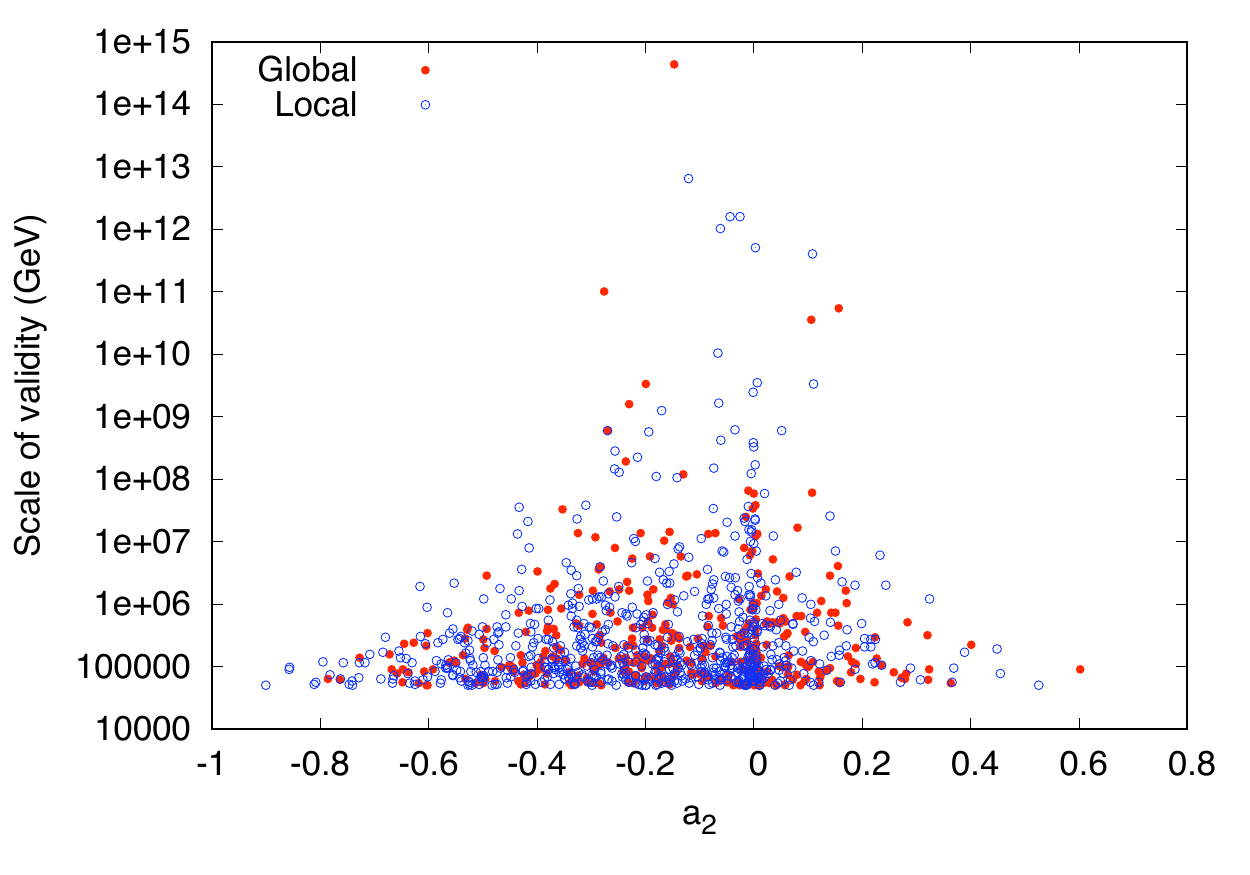}
\caption{\small Allowed region for $\lambda$ and $a_2$, where the EW vacuum is the ``global'' 
(red) or the ``local'' (blue) minimum in $Z_2$-asymmetric SM+S models, to have the validity at least upto 50 TeV.}
\label{fig:noz2val}
\end{figure}

In this section, we focus on only those models that allow two non-zero minima for the potential,
as discussed in Section~2.2. 
From \autoref{pot-min3-vs} we can see that for $b_3\ne0$ the two minima have unequal depths. 
We demand one of them to be the EW vacuum with 
$|v|=246$ GeV. If this is the deeper minimum, the universe is stable; if this is the shallower one, the universe 
can decay to the true vacuum and in that case the lifetime must at least be equal to the age of the universe. 

First, let us focus on the tree-level potential. In \autoref{fig:noz2val}, we show the range of validity 
of these models for various choices of $a_2$ and $\lambda$. The trend is similar to what we have seen before 
in Section 4.1: 
large values of $\lambda$, $a_2$, or even $b_4$ make the couplings blow up at a relatively low scale. Note that 
$a_2 < 0$ ($a_2 > 0$) models tend to have a local (global) minimum at $v=246$ GeV, but there are exceptions. 
The range of validity of these models is almost the same as that of the $Z_2$-symmetric case, 
because the RG evolutions of the couplings are controlled by the dimensionless couplings. 
The allowed region for $a_2$, however, is bunched more towards $a_2\sim 0$. 

\begin{figure}[t!]
\centering
\includegraphics[width=8.5cm]{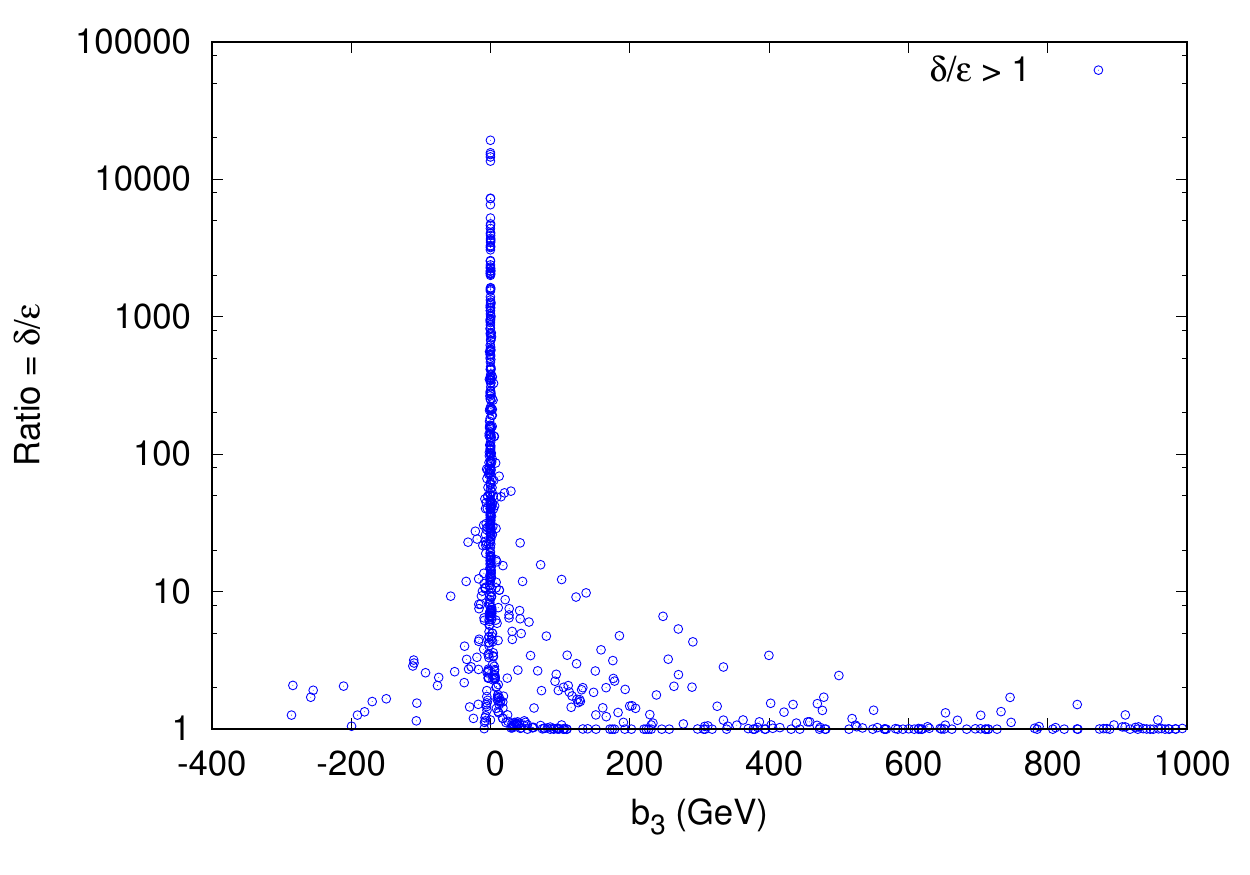}
\hspace{-0.35cm}
\includegraphics[width=8.5cm]{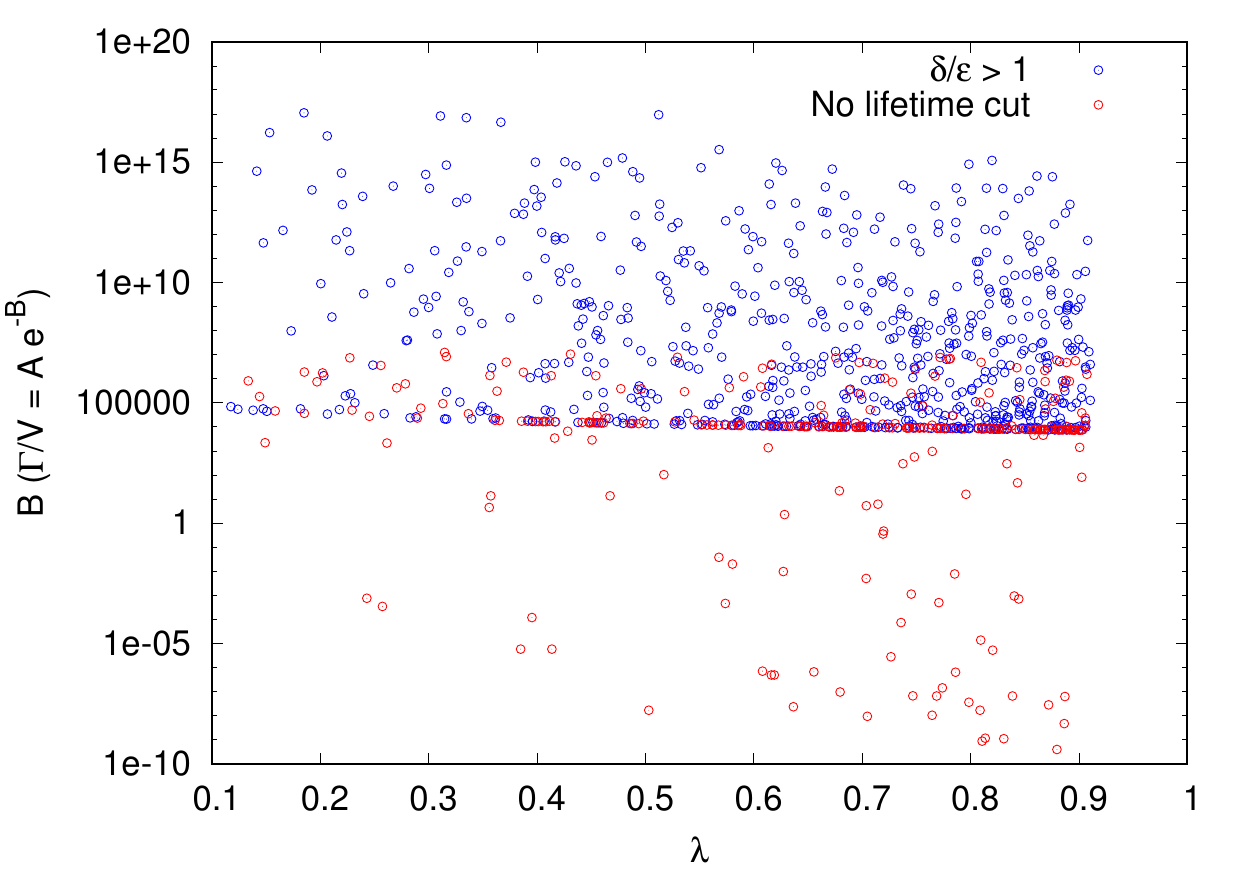}
\caption{\small The dependence of the lifetime of the metastable vacuum on
$b_3$ (left) and $\lambda$ (right).}
\label{fig:stable}
\end{figure}

For models with a shallower minimum at the EW vacuum, one may also estimate the lifetime of the universe. 
This is bound to be a rough estimate as the path between the two minima need not pass through a local 
maximum or even a saddle point. Approximately, $B \geq 1$ (see \autoref{B-def}) 
leads to a metastable vacuum \cite{barroso} 
while $B \leq 1$ tends to make the universe unstable. Assuming the maximum of the quartic couplings to 
be of the order of unity, this results in an approximate bound of $\delta/\epsilon > 0.05$.  

In \autoref{fig:stable} we show how the tunnelling lifetime depends on the parameter $b_3$ as a function of the 
ratio $\delta/\epsilon$ and $\lambda$ as a function of $B$. The dependence of $b_4$ is similar to that of $\lambda$.
\autoref{fig:stable} shows that the controlling parameter is $b_3$ because that creates the depth difference 
between the two minima. The smaller $b_3$ is, the larger is the lifetime of the metastable minimum.
As a conservative estimate, we have taken the stability limit to be $\delta/\epsilon \geq 1$, 
which excludes the low $\delta/\epsilon$ and hence low $B$ 
values. Such excluded models are also shown in the right hand side plot of \autoref{fig:stable}. 
The distribution of models does not depend much on the quartic couplings 
$b_4$ or $\lambda$.

\section{One loop corrections to the SM+S potential}
\label{oneloop}
\begin{figure}[t!]
\begin{center} 
\includegraphics[width=8.5cm]{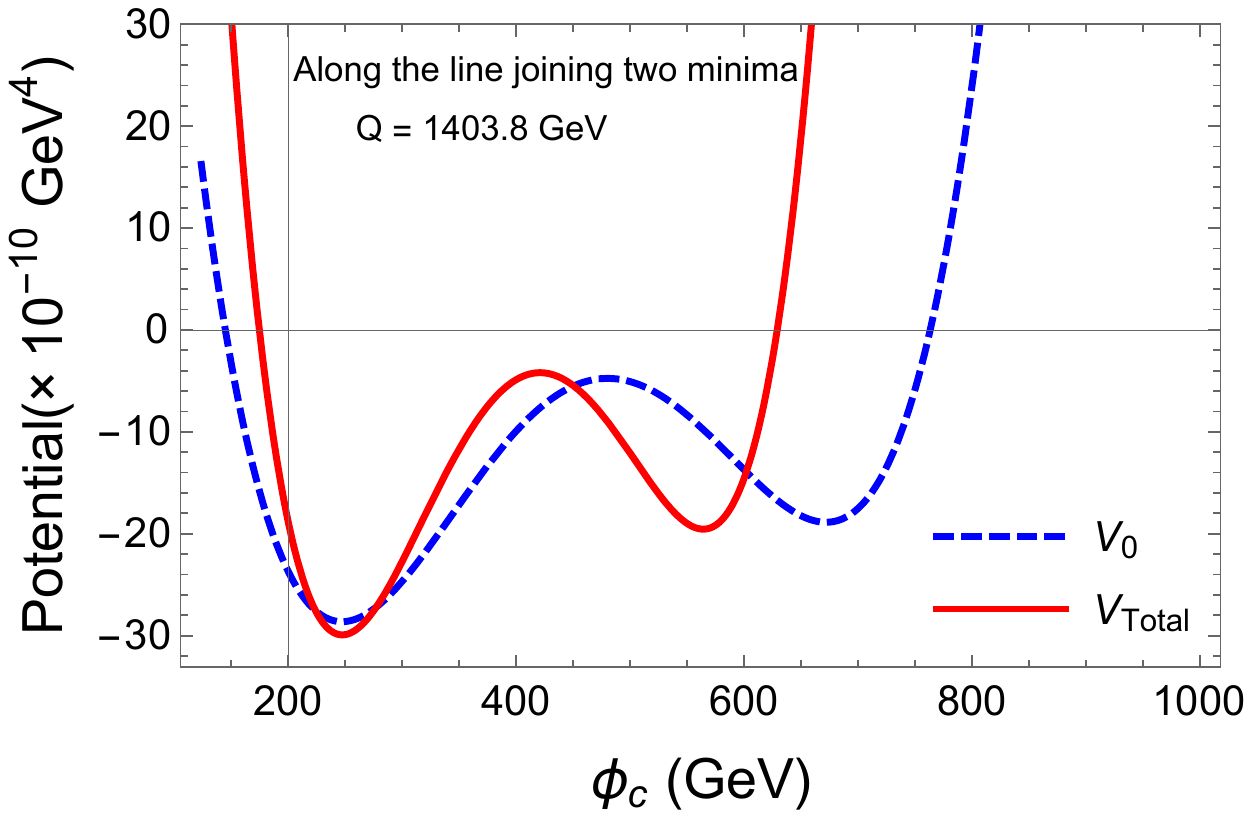}
\includegraphics[width=8.5cm]{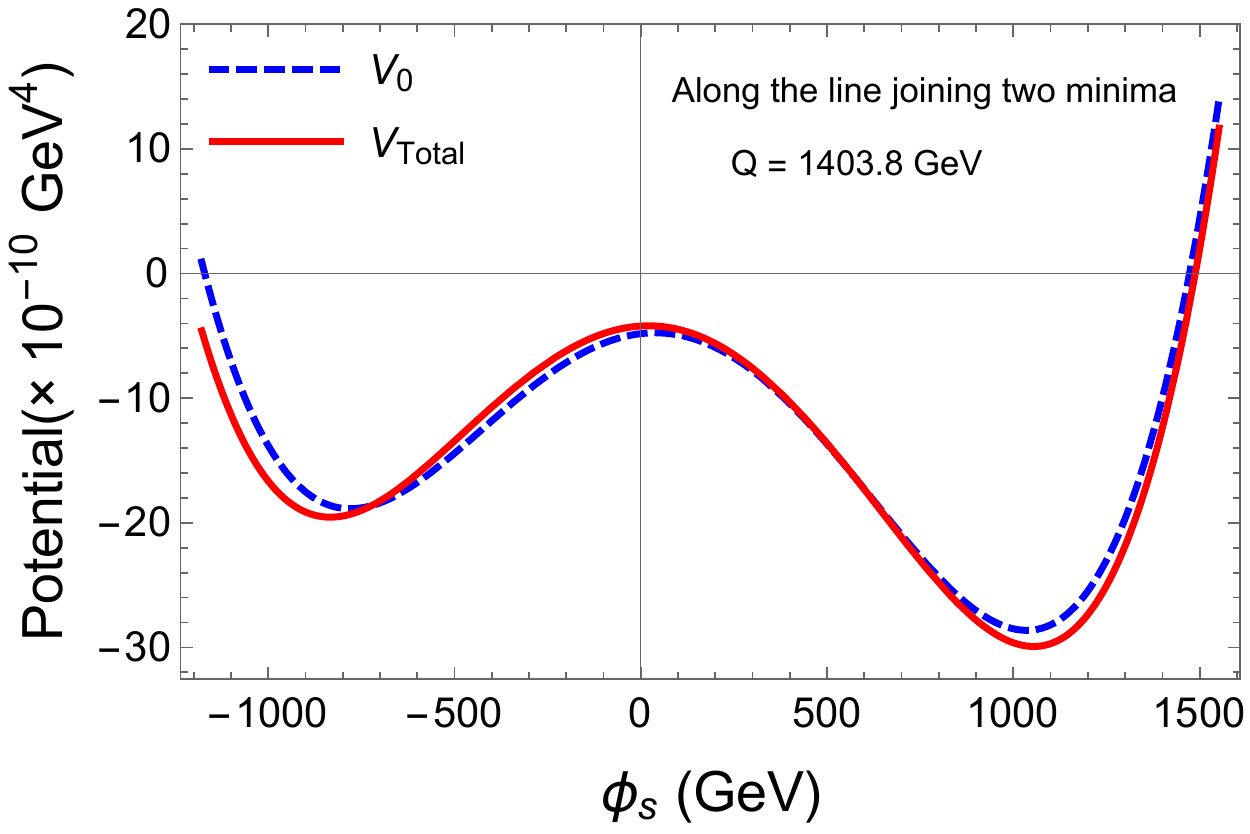}\\
\includegraphics[width=8.5cm]{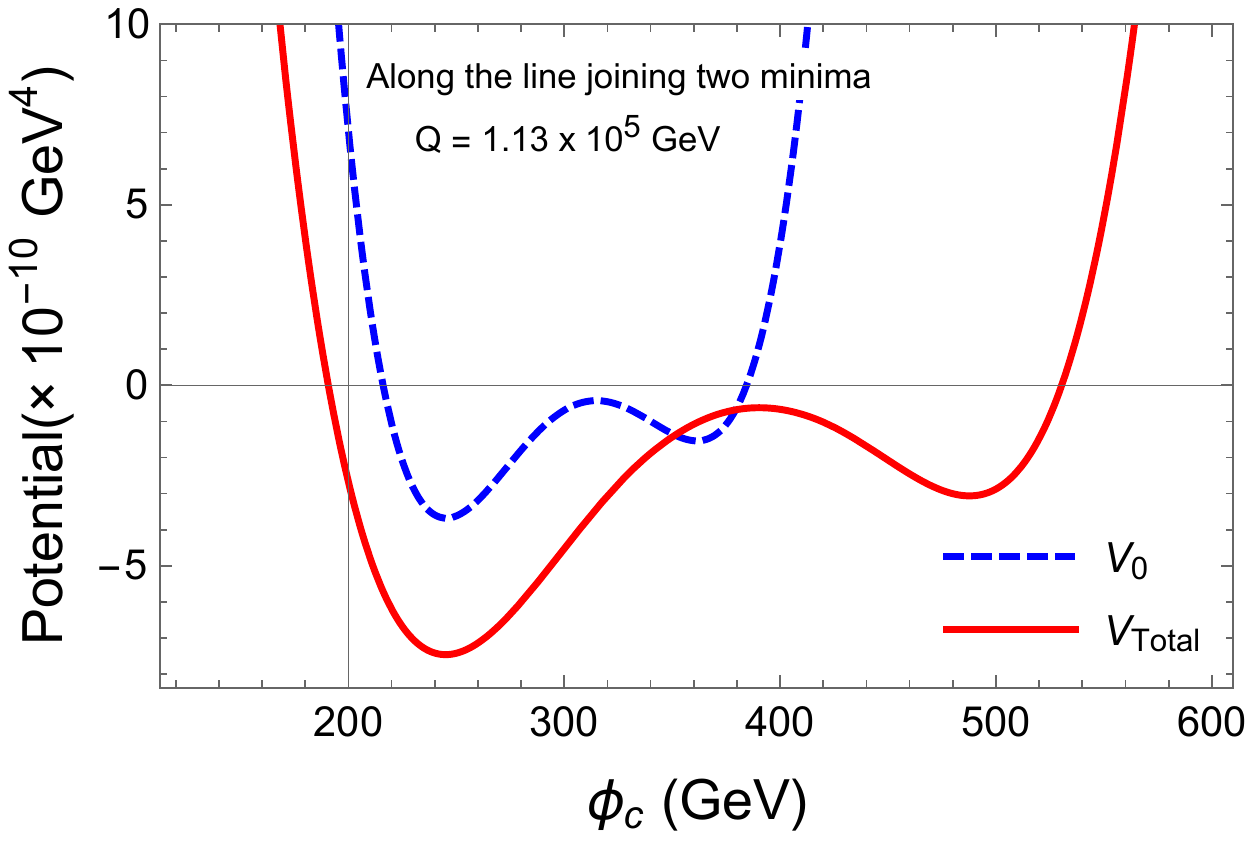}
\includegraphics[width=8.5cm]{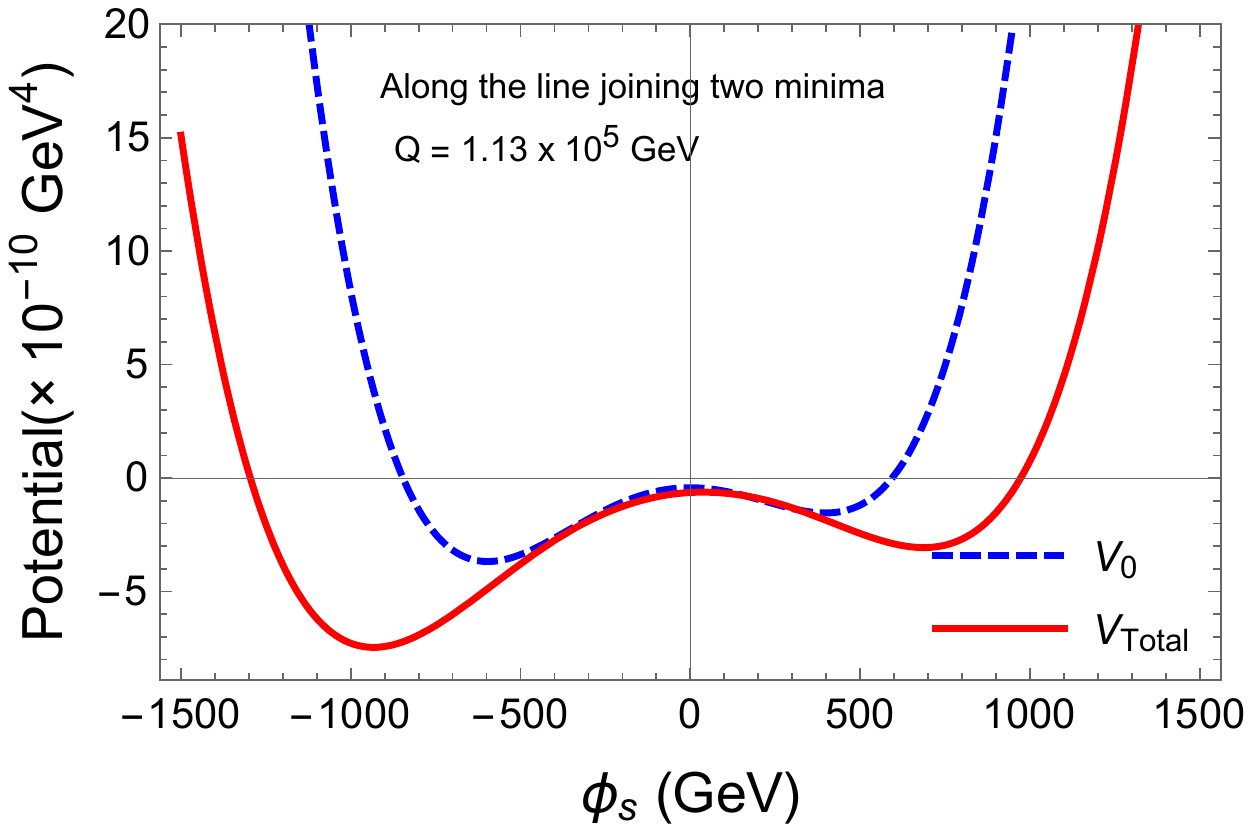}

\end{center}
\caption{\small The potential profile as a function of $\phi_c$ (the classical minimum for the doublet-dominated field) 
and $\phi_s$ (the classical minimum for the singlet-dominated field) for the two models described in the text. The upper panel 
is for model 1 and the lower panel is for model 2. The profiles are drawn along the line joining 
the two minima. Note that the second minimum can change significantly and the shift depends on the choice of the 
regularization scale $Q$. 
}
\label{fig:oneloop}
\end{figure}

The effect of one-loop corrections on the tree-level potential is not very drastic. This is expected because 
both VEVs are nonzero and neither $\Phi$ nor $S$ direction is a flat one. The one-loop corrections, being 
perturbative in nature, are suppressed by the standard loop factor of $1/64\pi^2$ and are 
expected to be significant only if we look at a flat direction. 

We show the one-loop corrections for two models, the parameters at the electroweak scale are 
given in \autoref{table}.
\begin{table}
\begin{center}
\begin{tabular}{|l|c|c|}
\hline
 & Model 1 & Model 2 \\
 \hline
 $a_1$ (GeV)  & $-20.8$ & $90.1$ \\
 $a_2$ & $0.40$ & $0.22$ \\
 $b_3$ (GeV)  & $-58.6$ & $79.3$\\
 $b_4$ & $0.26$ & $0.33$\\
 $\lambda$ & $0.46$ & $0.60$\\
 $\mu^2$ (GeV$^2$) & $4.58\times 10^5$ & $1.16\times 10^5$ \\
 $M^2$ (GeV$^2$) & $4.85\times 10^5$ & $1.72\times 10^5$ \\ 
 Global min.: $(v, v_s)$ (GeV) & $(247, 1035)$ & $(245, -598)$\\
 Local min.: $(v, v_s)$ (GeV) & $(673,-786)$ & $(362, 406)$ \\
 \hline
 \end{tabular}
 \end{center}
 \caption{Parameter values for the models for which the 1-loop corrections 
 are shown in the \autoref{fig:oneloop}.}
 \label{table}
\end{table}
We choose the 
regularization scale $Q$ in such a way that even after the one-loop corrections, the EW vacuum stays at 
$v \approx 246$ GeV (so that all SM particles have their masses unaffected). This choice of regularization scale was
motivated in Ref.\ \cite{indrani-4}. We show the effect of one-loop corrections in \autoref{fig:oneloop}
for these two models. While they both have the global minimum at $v \approx246$ GeV, the required regularization scales differ by more than one order of magnitude. The nature of change is similar for all 
models, and thus we do not expect an unstable vacuum model to become metastable (or vice versa) 
because of the one-loop corrections. However, one may note how much the global minimum has been lowered by the 
one-loop corrections for the second model. In fact, for all the models scanned, we have never found a switch from 
global to local minimum induced by the radiative corrections. 

\section{Summary}
\label{summary}

The potential of SM+S, a real singlet enhanced SM, shows several interesting features. In this paper, we 
have investigated the parameter space for several types of SM+S: the Higgs portal dark matter models, the 
potential with an explicit $Z_2$-symmetry and having singlet-doublet mixing, the $Z_2$-asymmetric potential 
with a stable EW minimum, or the same with an unstable or metastable EW minimum. The general features can be 
summarized as follows.

\begin{itemize}
 \item 
 Adding one more real singlet makes the potential less stable in general at a high energy. This happens 
 because the renormalization group equations for the couplings tend to hit the Landau pole much below the 
 Planck scale, more so if the starting values at the EW scale is large. All the quartic couplings, namely, 
 $\lambda$, $b_4$, and $a_2$, have to be small at the EW scale to keep the model valid up to a high scale, 
 as the $\beta$-functions are coupled. (However, such new scalar couplings are helpful to avoid the vacuum stability 
 bound, coming from the negative pull caused by the large top Yukawa coupling. Again, such conclusions 
 are not valid if there are more degrees of freedom, like vectorlike fermions.)
 Our numerical results are shown at one-loop but inclusion of higher-order
 corrections do not change the result qualitatively. In particular, if we start with a large value of 
 either $\lambda$ or $b_4$, the model hits the Landau pole at a relatively low scale. This can be taken as a possible 
 indication of some new physics taking over and one should consider the effect of higher-dimensional operators on the
 low-scale physics.
 \item
 While there are some minor variations for the allowed range of parameters among different class of models ({\em e.g.}
 $a_2$), the overlap is significant, and so one has to determine all the couplings experimentally to know what 
 class of SM+S it really is. This is, of course, an extremely challenging task, if not outright impossible; the 
 determination of $b_4$ is apparently beyond the reach of any present or upcoming colliders unless there is a 
 significant mixing between the singlet and the doublet and the Higgs self-couplings are determined with sufficient 
 accuracy. For the prospect of determination of the singlet-doublet mixing angle in future colliders, we refer the reader 
 to Ref.\ \cite{1407.5342}.
 \item
 The tree-level results are robust enough as far as the metastability issue is concerned. This is expected as we 
 are not looking along any flat direction. However, the one-loop corrections can change the position of the minima. 
\end{itemize}

\centerline{\bf{Acknowledgements}}

We thank Filippo Sala for pointing out a mistake in the first version of the paper. 
S.G. acknowledges the University Grants Commission, Government of India, for a research fellowship. A.K. 
acknowledges Department of Science and Technology, Government of India, and Council for Scientific and 
Industrial Research, Government of India, for extramural projects. 
S.R. likes to thank the Department of Science and Technology (DST),
Government of India for the INSA-INSPIRE Faculty
Fellowship. S.R. would also like to thank Prof. Yuval Grossman, LEPP, Cornell University for hosting 
the visit as a part of INSA-INSPIRE Faculty Fellowship, where a part of this work has been done.


\end{document}